\documentclass[pra,twocolumn,aps]{revtex4}
\usepackage{graphicx}  
\usepackage{dcolumn}   
\usepackage{bm}        
\usepackage{amssymb}   
\usepackage{amsmath}
\usepackage{setspace} 

\usepackage{hyperref}
\hypersetup{
  colorlinks   = true,    
  urlcolor     = blue,    
  linkcolor    = blue,    
  citecolor    = red      
}

\begin{document}

\title{Perturbative approach  in the frequency domain for the intensity correlation spectrum in electromagnetically induced transparency 
}

\author{H. M. Florez $^{1}$, C. Gonz\'alez$^{1}$, and M. Martinelli$^{1}$}

\affiliation{$^{1}$Instituto de F\'{\i}sica, Universidade de S\~ao Paulo, 05315-970 S\~ao Paulo, SP-Brazil}

\begin{abstract} 
Correlation spectroscopy has been proposed as a spectroscopic technique for measuring the coherence between the ground states 
in electromagnetically induced transparency (EIT). 
While in the time domain the steep dispersion in EIT condition accounts for the robustness of the correlation linewidth against power broadening,  such physical insight was not directly established in the frequency domain. 
We propose a perturbative approach to describe the correlation spectroscopy of two noisy lasers coupled to a $\Lambda$-transition in cold atoms, leading to  EIT.
  Such approach leads to an analytical expression that maps the intensity correlation directly in terms of the absorption and dispersion of the light fields. Low and high perturbative  regimes are investigated and demonstrate that, for coherent light sources, the first oder term in perturbation expansion represents a sufficient 
  description for the correlation. Sideband resonances are also observed
 , showing the richness of the frequency domain approach.
\end{abstract}

\maketitle



 
The coherent interaction between light and matter has been a subject of fundamental interest in quantum communication and quantum information \cite{Lukin03,Polzik10}. Such coherence opens the possibility of mapping information between light and matter, which is necessary  for developing quantum memories and quantum repeaters \cite{Marangos05,Duan2001,Riedmatten11}. 
Among those coherent interactions, electromagnetically induced transparency (EIT) has been spectroscopically studied 
\cite{Harris,HarrisExp}.
The EIT phenomenon in a $\Lambda$ transition relies on the interference between atomic excitations driven by two light fields  pumping the atomic state into a superposition between ground states, known as a dark state.
In this dark state the medium becomes transparent to light when the optical frequencies satisfy the two photon resonance, with a linewidth narrower than the natural linewidth.
Given this sub-natural character of the EIT linewidth, 
it finds applications in atomic clock stabilization  and cold atom thermometry~\cite{Vanier05, Halfmann12}. 
Nevertheless, the linewidth of the EIT signal
is sensitive not only to the decoherence between the ground states but also to power broadening.

Usually,  the EIT linewidth is measured by transmission spectroscopy, with the measurement of the mean value of the transmitted intensity.
On the other hand,  the sensitivity increases when we look at the intensity fluctuations of the field  when noisy light sources   are employed for the spectroscopy. Yabuzaki \textit{et al.}  \cite{Yabusaki91} showed that this type of spectroscopy is characterized by the transformation of excess phase noise into amplitude noise, due to the interaction with a two-level atom. Moreover, instead of only measuring the intensity  noise of the light beams, measuring the intensity correlation of two beams interacting in EIT condition has provided new 
insights~\cite{Garrido03,Martinelli04,Cruz07,Scully05,Scully08,Scully10,Xiao09}.  
Recent contributions~\cite{Xiao09,Felinto13} showed that, by performing intensity correlation spectroscopy with noisy lasers, it is possible to measure the intrinsic EIT linewidth, which is narrower than the EIT linewidth measured by standard transmission spectroscopy.  In particular, in Ref.~\cite{Felinto13} a simple model was developed that allows to connect the intrinsic EIT linewidth with the decoherence that limits the lifetime of the ground states' superposition, for cold atoms. This technique can work as a tool to directly measure the decoherence between ground states, since it is free from power broadening. 

Although the intensity correlation was  well described by numerical calculations, the lack of an analytical model prevents a deeper physical 
insight  that would help to establish which atomic properties determine the spectroscopic features of intensity correlation. The physical description of the correlation can
be done either in the time domain by the  normalized two time correlation function $g^2(\tau)$ or in the frequency domain by a normalized version of its Fourier transform, $C(\omega)$. In the time domain,  Felinto \textit{et al}.~\cite{Felinto13} proposed a simple model for $g^2(0)$ 
that illustrates the role of the steep dispersion in preventing the power broadening of the intrinsic linewidth.
The numerical evaluation of the correlation in the frequency domain is successful in showing the same feature, as it was shown in refs.\cite{Cruz07,HMFlorez13}. 
However, an analytical description was still missing. 

In this paper  we propose a perturbative approach to determine the intensity correlation between two light fields in EIT with cold atoms, when described in the frequency domain. The perturbative expansion lets us  identify the main atomic contributions to the intensity correlation between the beams, in both time and frequency domains. From the perturbative approach we obtained an analytical expression for the correlation coefficient. 

We  show that the atomic response to a low noise laser can be fully described by the first order term in the expansion. 
This 
first order term is determined by the absorption and dispersion of the fields, as in the heuristic 
model in ref.~\cite{Felinto13}. For noisy laser sources,  higher order terms in the perturbative expansion are considered,
recovering the correlation profile obtained in refs.~\cite{Xiao09,Felinto13,HMFlorez13}.  We also show that the correlation coefficient presents the contribution
from the resonances of the sidebands of the laser, shifted from the central carrier by the analysis frequency $\omega$.
Their contribution results in a broadening of the correlation profile inside the transparency window. 
Nevertheless, the intrinsic linewidth of the correlation peak is completely invariant to the analysis frequency, showing that 
 both approaches are equivalent for measuring the coherence lifetime of the ground states.

The paper is organized as follows. In section \ref{sec:corr}, we briefly describe the atomic levels for  studying the  correlation spectroscopy. We also introduce  the approach of stochastic variables used in the calculation of the covariance matrix associated to the atomic density operator and the corresponding spectral density matrix. In section \ref{sec:PertModel}, we present the perturbative expansion to determine the $g^2(0)$ function and  the correlation coefficient $C(\omega)$.  In section \ref{sec:Mapping}, we decompose the perturbative result to obtain the analytical expression of the correlation in terms of absorption and dispersion of the fields. In section \ref{sec:sidebands} we show the role of sideband resonances in the correlation spectroscopy and in section \ref{sec:conclusions}, we summarize the results of the perturbative approach for correlation spectroscopy.

\section{Correlation spectroscopy}\label{sec:corr}

The correlation between any two light fields with intensities $I_1(t)$ and $I_2(t+\tau)$ can be quantified by the $g^2(\tau)$ function \cite{Scully05}. 
In particular, we study the intensity correlation between two noisy lasers (a diode laser for example) induced by cold atomic media in EIT condition.
The use of noisy lasers  limits the intensity correlation to those obtained by classical states of light (i.e, those described by regular Glauber-Sudarshan P-functions~\cite{GlauberSudarshan}).
Yabuzaki \textit{et al}. showed in ref.\cite{Yabusaki91} that the atomic medium transforms excess of phase noise of the input light sources into intensity noise at the output.
Hence,  the phase  noise to amplitude noise (PN-AN) transformation has to be considered for the theoretical description of the intensity correlation in our bipartite system in EIT condition. 
Here we will present a semi-classical approach, where the atomic response is treated by quantum mechanics while the light fields are considered to be classical, with  stochastic phase fluctuations. Therefore, 
intensity fluctuations at the input are not considered. 

Let us consider two electromagnetic fields described by 
\begin{equation}
\mathbf{E}_{i}(t)={\cal E}_{i} \exp \left[i(\omega_{i} t+ \phi_{i}(t))\right] \mathbf{e}_{i},
\end{equation}
with a stochastic phase fluctuation $ \phi_{i}$ that models  the excess of noise in diode lasers.
In the expression for the fields, $i=1, 2$ denotes the two beams, ${\cal E}_{i}$ and $\omega_{i}$ are respectively the amplitude and the frequency, and $\mathbf{e}_{i}$ stands for the distinct polarization modes.

The converted phase noise is easily measurable  by photodetectors.
This can be shown if we consider that after the propagation of the field through the medium,  the output field in the thin sample limit~\cite{Zoller94} is given by
\begin{equation}\mathbf{E}'_i(t) = \mathbf{E}_i(t) + i \kappa\mathbf{P}_i(t),
\end{equation}
where polarization $\mathbf{P}_i(t)$ represents the atomic response induced by the incident fields $\mathbf{E}_i(t)$ and $\kappa$ is a real constant that depends  on the atomic density. 

\begin{figure}[htb!]
\centering
\includegraphics[width=86mm]{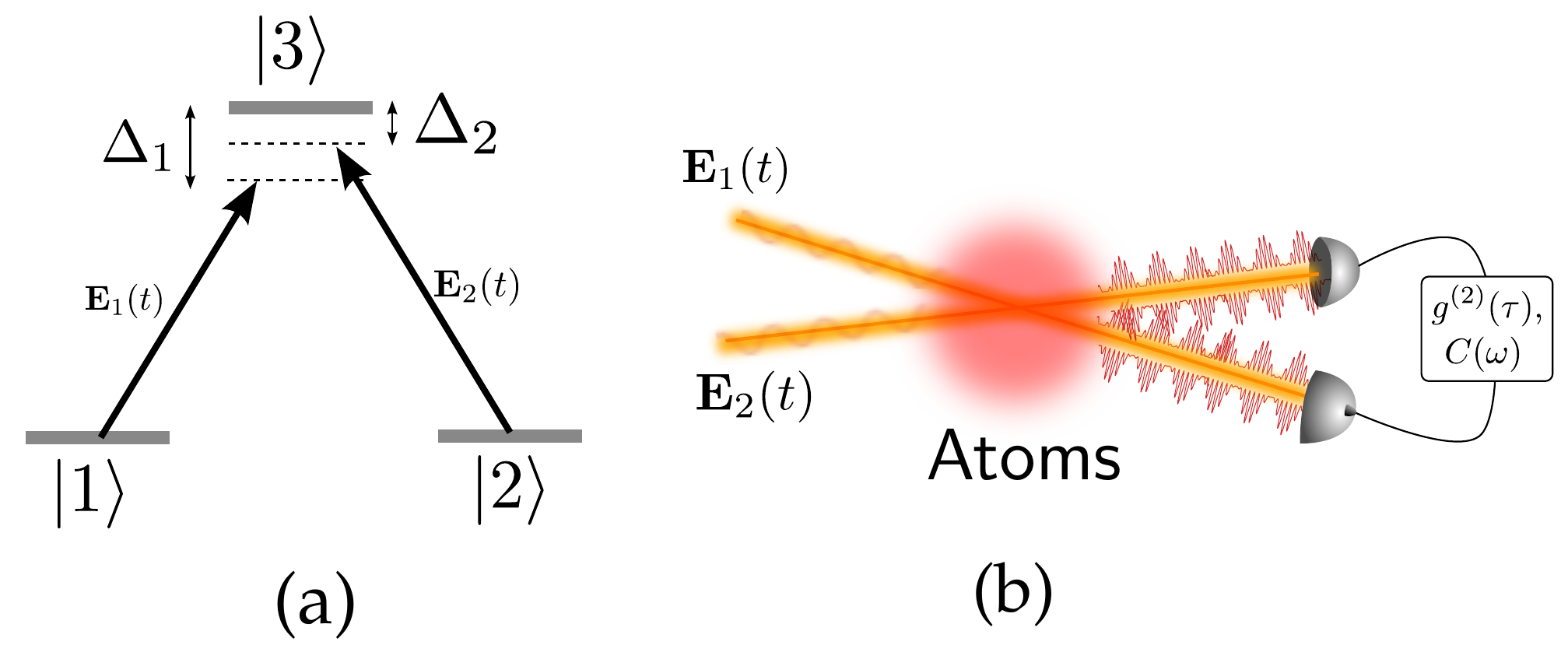}
\vspace{-.5cm}
\caption{(Color online) (a) Levels scheme in $\Lambda$-type configuration for EIT. (b) Basic setup for measuring the intensity correlation spectrum in time ($g^2(\tau)$) or frequency domain ($C(\omega)$).}
\label{fig:Levels}
\end{figure}

The usual scheme for correlation spectroscopy in a $\Lambda$- EIT condition is shown in Fig.~\ref{fig:Levels}(a).  Two fields,  $\mathbf{E}_{1}(t)$ and  $\mathbf{E}_{2}(t)$ are coupled to two different transitions of a three-level atom, with different one-photon detunings $\Delta_1$ and $\Delta_2$ respectively. The intensity correlation between the two fields is measured after interacting with the atomic ensemble as presented in Fig.\ref{fig:Levels}(b). 
 
The incident fields are coupled through the $\Lambda$ transition,  inducing  polarizations  of the atomic medium that contribute to each output field as $\mathbf{P}_i= \mathbf{d}_{i3} \rho_{i3}$, where $\mathbf{d}_{i3}$ and $\rho_{i3}$ represent the electric dipole moment and the atomic coherence associated to the fields $i=1,2$. The  output intensity of each field is then given by 
\begin{equation}
I'_{i}(t)=I_{i} - 2\varrho_i\mathcal{E}_i\text{Im} \{\rho_{i3}e^{i\phi_i}\}, \label{intensity}
\end{equation}
where $\varrho_i=\kappa \mathbf{d}_i\cdot\mathbf{e}_i$ and the intensity is expressed up to first order in $\kappa$.
The intensity correlation between two light fields is defined as 
\begin{equation}
g^2(\tau) = \frac{\langle \delta I'_1 (t ) \delta I'_2 (t + \tau)\rangle  }{\sqrt{\langle  \delta I'_1 (t)^2\rangle \langle \delta I'_2 (t + \tau)^2\rangle}},
\label{eq:cor}
\end{equation}
where $\delta I'_1(t)$ and $\delta I'_2(t+\tau)$ represent the light intensity fluctuations of each beam at different times separated by a time delay $\tau$. For $\tau\sim 0$, 
the atomic response induces correlation ($g^2(0)>0$) or anti-correlation ($g^2(0)<0$) between the light fields~\cite{Scully05,Xiao09,Felinto13,HMFlorez13}, depending on the two-photon detuning $\delta=\Delta_1-\Delta_2$.

The heuristic model 
proposed in ref.~\cite{Felinto13} establishes a direct relation between the atomic response and the $g^2(0)$ function.  It was shown  that the stochastic phase fluctuations were not only mapped by the absorptive properties of the medium, but also by the dispersive atomic response. In the thin sample limit, the intensity fluctuations are proportional to the atomic coherences, i.e. $\delta I'_i(t)\sim \delta \text{Im}\{p_i e^{i\phi_i}\}$, where $p_i= \rho_{i3}^{ss}$ is the stationary solution of the density matrix element contributing to the intensity fluctuation in eq.~(\ref{intensity}).
As a consequence of the time averaging of the phase diffusion,  $\delta I'_i(t)$ can be expressed in terms of  $\text{Re}\ p_i$ and $\text{Im}\ p_i$, and therefore,  the correlation function $g^2(0)$ is also written in terms of the contributions of atomic dispersion ($\text{Re}\ p_i$) and absorption ($\text{Im}\ p_i$) for each beam as 
\begin{equation}
g^2(0)=\frac{\text{Re}\ p_1 \ \text{Re}\ p_2 +\text{Im}\ p_1 \ \text{Im}\ p_2 }{\sqrt{(\text{Re}^2 p_1+\text{Im}^2 p_1)(\text{Re}^2 p_2+\text{Im}^2 p_2)}}. \label{g20}
\end{equation}

It was shown that in the low power regime, the absorption dominates and the medium induces correlation between both light fields. As the power of the fields is increased, the dispersive term  $\text{Re}\ p_1\ \text{Re}\ p_2<0$ overcomes the contribution from the absorptive term $\text{Im}\ p_1\ \text{Im}\ p_2>0$, leading to anti-correlated light fields i.e. $g^2(0)<0$. 
 
Moving to the  frequency domain, the noise correlation is defined as the normalized Fourier transform of the correlation function in Eq. (\ref{eq:cor}), that can be described as
 \begin{equation}
C(\omega) = \frac{S^I_{12} (\omega)}{\sqrt{S^I_{11}(\omega)S^I_{22}(\omega)} },
\label{eq:corfreq}
\end{equation}
where $S^I_{ij} (\omega)$ represents the symmetrical intensity correlation spectrum for  $i$ and $j$ fields at a given analysis frequency $\omega$ such that
\begin{align}
 S^I_{ij}(\omega)  =&\frac{1}{2\pi}\int_{-\infty}^\infty d\tau e^{-i\omega\tau} \langle I'_i(t) , I'_j(t+\tau) \rangle_{S}.
 \label{SpectrInt}
 \end{align}
where we define the symmetrical covariance $\langle I'_i(t) , I'_j(t+\tau) \rangle_{S}=(\langle I'_i(t) , I'_j(t+\tau) \rangle+\langle I'_j(t) , I'_i(t+\tau) \rangle)/2$ with 
$\langle \Upsilon_i,\Upsilon_j\rangle=\langle \Upsilon_i\Upsilon_j\rangle-\langle \Upsilon_i\rangle \langle\Upsilon_j\rangle$ as the covariance for any stochastic variable $\Upsilon_i$.
Similar to $g^2(0)$, 
for correlated fields  $C(\omega)>0$, and for anti-correlated fields $C(\omega)<0$. 

The theoretical approach used in the evaluation of this correlation adopted in refs.~\cite{Martinelli04,Cruz07,HMFlorez13}, 
provides a result that is consistent with experimental data. Nevertheless, they do not present the detailed role of each term 
of atomic absorption and dispersion, as was done in ref.~\cite{Felinto13} in evaluation of eq.~(\ref{g20}). In order to overcome 
this limitation, we study the stochastic dynamic for the $\Lambda$-EIT system using Ito's calculus, 
which is the platform for the perturbative expansion performed in section \ref{sec:PertModel}.

\subsection{Langevin equation for the atomic dynamics from Ito's calculus}

Following the theoretical approach in refs. \cite{Cruz07,HMFlorez13}, let us consider a three level system interacting with two
light fields $E_1$ and $E_2$ in  a $\Lambda$  configuration as shown in Fig. \ref{fig:Levels}(a). The interaction hamiltonian is given by
\begin{equation}
H_{\text{int}}=\hbar \left( \Omega_1^* 
\hat{\sigma}_{13}e^{i (\omega_1 t + \phi_1(t))} +\Omega_2^*\hat{\sigma}_{23}e^{i (\omega_2 t +\phi_2(t))} + \text{h.c.}\right),  \label{hamiltonian}
\end{equation}
where $\hat{\sigma}_{i3}=|i\rangle \langle3|$ are the atomic operators associated with each transition, 
$\Omega_i$ are the Rabi frequencies, $\omega_i$ are the optical frequencies and
$\phi_i(t)$ are the stochastic phase fluctuations, associated to the fields $\mathcal{E}_i$ for $i=1,2$. The time evolution of the phase fluctuations $\phi_i(t)$ is described by a Wiener process, therefore their statistical moments satisfy
\begin{subequations}
\begin{align}
\langle  d\phi_i(t)\rangle=&0,\label{dwmedia}\\
   \langle d\phi_i(t),d\phi_i(t)\rangle =& 2\gamma_{i}dt,\label{dw2media}\\
 \langle d\phi_1(t),d\phi_2(t)\rangle =& 2\gamma_{12} dt,\label{dw2mediaX}\\
   \langle d\phi_i(t)^n, d\phi_j(t)^m\rangle=&\ 0 \hspace{.2cm} \text{para\ } n\geq2\ ,m\geq 2 ,
\end{align}\label{dphi}
\end{subequations}
where $\langle \cdots\rangle$  denotes stochastic average,  $\langle \cdots, \cdots\rangle$ stands for the covariance and $\gamma_i$ represents the linewidth of the Lorentzian lineshape that spectrally  characterizes the laser's phase fluctuations.  Considering the experimental conditions where the two beams are generated
by the same laser source, we can assume the same linewidth for both beams, with correlated fluctuations, such that
$\gamma_1=\gamma_2=\gamma_{12}=\bar{\gamma}$.

The atomic dynamics is described by the Bloch equations for the atomic matrix elements $\tilde{\rho}_{ij}$, which are obtained from the interaction hamiltonian of the system. Their evolution is better described with the use of Liouville space. 
In particular for our three level system, we use the vector   
$\mathbf{y}=(\tilde{\rho}_{11},\tilde{\rho}_{22},\tilde{\rho}_{13},\tilde{\rho}_{31},\tilde{\rho}_{23},\tilde{\rho}_{32},\tilde{\rho}_{12},\tilde{\rho}_{21})$ 
to describe all the independent components. The Bloch equations are determined in Appendix \ref{apendBloch} and they describe 
the dynamics for the rapidly varying variables $\mathbf{y}$. To obtain a linear system without explicit time dependence,
the atomic dynamics is better described by the slowly varying variables
$\mathbf{x}=(\rho_{11},\rho_{22},\rho_{13},\rho_{31},\rho_{23},\rho_{32},\rho_{12},\rho_{21})$. 
Hence, it is convenient to perform the rapidly-to-slowly varying variables transformation such that 
$\tilde{\rho}_{i3}e^{-i(\omega_i t+\phi_i(t))}\rightarrow\rho_{i3}$ and $\tilde{\rho}_{12}e^{-i((\omega_1-\omega_2) t+\phi_1(t)-\omega_2(t))}\rightarrow\rho_{12}$.
Such transformation is represented in a very compact way as
 \begin{align}
 \mathbf{x}(t)=e^{-i\textbf{N}_1(\omega_1 t+\phi_1(t))}e^{-i\textbf{N}_2(\omega_2 t+\phi_2(t))}  \mathbf{y}(t) , \label{yxtransform}
\end{align}
where $\mathbf{N}_1$ and $\mathbf{N}_2$ are diagonal matrices with zeros and ones 
defined in Appendix \ref{apendBloch}. Each one of the matrices $\mathbf{N}_i$ corresponds to each noisy beam $i=1,2$.
Therefore, the matrix representation of the Bloch equations is given by  
\begin{align}
\frac{d\mathbf{y}(t)}{dt}=-e^{i\mathbf{N}_1(\omega_1 t+\phi_1(t))}e^{i\mathbf{N}_2(\omega_2 t+\phi_2(t))} \mathcal{A} \mathbf{x}(t) + \mathbf{y}_0, \label{dydt}
\end{align}
where the $\mathbf{y}_0$ vector and the $\mathcal{A}$ matrix are both defined explicitly for a three level system in Appendix 
\ref{apendBloch}. The vector $\mathbf{y}_0$ is constant, used for normalization of the closed three-level system.
The evolution matrix $\mathcal{A}$ depends on parameters of the light-atom interaction such as the spontaneous decay rate $\Gamma$,
the ground state coherence lifetime $\gamma_d$, the Rabi frequencies $\Omega_i$ for each field mode and the resonant frequencies
for the atomic transitions $\omega_{(13)}$ and $\omega_{(23)}$. 
It is worth noting that the matrix representation in eq.(\ref{dydt}) stands for any atomic structure (with $M$ levels) that couples two noisy beams and additional monochromatic noiseless fields.
In that general case, the only difference would be the higher dimension to be considered for the vectors and matrices 
in eq.(\ref{dydt}), such as the five level system in ref. \cite{Xiao14_2}. Even more general, if we consider a M-level 
atom interacting with more than two noisy beams, the only difference, besides higher dimensions, are the extra terms for 
$\exp \{i\mathbf{N}_1(\omega_1 t+\phi_1(t))+\cdots +i\mathbf{N}_n(\omega_n t+\phi_n(t))\}$ in eq.(\ref{dydt}) where $n$ 
would be the total number of fields with stochastic phase noise. Thus, the theoretical treatment presented here
can be easily extended for M-level atomic systems interacting with two noisy fields. Nevertheless, we restrict ourselves here to describe a three level system.

Once we obtain the Bloch equations, now we determine the Langevin equations 
for the slowly varying variable $\mathbf{x}$ with the help of Ito's calculus \cite{Gardiner90} for the stochastic 
fluctuations $d\phi(t)$. By applying Ito's differentiation rule 
$d\mathbf{x}=[\partial_t \mathbf{x} + \frac{1}{2}(\partial_{\phi_1}^2 \mathbf{x}+\partial_{\phi_2}^2 \mathbf{x}
+\partial_{\phi_1\phi_2}^2 \mathbf{x}+\partial_{\phi_2\phi_1}^2 \mathbf{x})]dt +\partial_{\phi_1}\mathbf{x}\ d\phi_1  +
\partial_{\phi_2}\mathbf{x}\ d\phi_2 $ into the transformation (\ref{yxtransform}), we obtain a general Langevin equation
\begin{align}
d\mathbf{x}(t)&=\mathbf{a}[\mathbf{x},t]dt  +\mathbf{B_1}[\mathbf{x},t] d\phi_1(t)+ \mathbf{B_2}[\mathbf{x},t] d\phi_2(t)\nonumber\\
&- \frac{1}{2}( \mathbf{N}_1^2d\phi_1^2 +  \mathbf{N}_2^2d\phi_2^2 + 2 \mathbf{N}_1 \mathbf{N}_2d\phi_1d\phi_2)\mathbf{x}(t)
\label{langevin1}
\end{align}
with
\begin{subequations}
\begin{align}
 \mathbf{a}[\mathbf{x},t]&=-\mathbf{A}_t\ \mathbf{x}(t)+\mathbf{x}_0,\\
         \mathbf{B_1}[\mathbf{x},t]&=i\mathbf{N}_1\ \mathbf{x}(t),\\
         \mathbf{B_2}[\mathbf{x},t]&=i\mathbf{N}_2\ \mathbf{x}(t),
\end{align}
\end{subequations}	
where $\mathbf{A}_t=\mathcal{A}-i\mathbf{N}_1\omega_1-i\mathbf{N}_2\omega_2$, and
$\mathbf{x}_0$ corresponds to the transformation of $\mathbf{y}_0$, following eq.(\ref{yxtransform}). 
The new evolution matrix $\mathbf{A}_t$ has the same matrix form as $\mathcal{A}$ in eq.(\ref{MatrixA3N}), 
but in that case we change $\omega_{(3i)}\rightarrow \Delta_{i}$ with  $\Delta_i= \omega_i-\omega_{(3i)}$ 
which represent the optical frequency detunings with respect to the atomic resonances for both beams $i=1,2$ (see Fig.\ref{fig:Levels}). 

Notice that the atomic evolution given by eq. (\ref{langevin1}) is written  as a contribution of two main parts. The dynamics determined by noiseless fields is described by the vector $\mathbf{a}$, while the contribution of stochastic phase fluctuations $d\phi_i(t)$  into the atomic dynamics depends on $\mathbf{N}_i$ matrices up to second order. These phase fluctuations are amplified by the atomic variables due to the product $d\phi \times\mathbf{x}(t)$ and are fed back into the system evolution by $d\mathbf{x}(t)$.

According to the statistical properties of the phase fluctuations in eqs.(\ref{dphi}), the mean value of the atomic variables evolves slowly as
\begin{align}
d\langle\mathbf{x}(t)\rangle&=[-\mathbf{M} \langle \mathbf{x}(t)\rangle + \mathbf{x}_o]dt
\label{dxdt}
\end{align}
with
 $\mathbf{M}=\mathbf{A}_t+\bar{\gamma}[ \mathbf{N}_1^2 + \mathbf{N}_2^2+2 \mathbf{N}_1  \mathbf{N}_2]$, 
 reaching the stationary state of the mean value given by
\begin{align}
\langle\mathbf{x}(t)\rangle_{ss}&=\mathbf{M}^{-1}  \mathbf{x}_o.
\label{x_ss}
\end{align}
The components $\rho_{13}^{ss}$ and $\rho_{23}^{ss}$ correspond to  the stationary solution for the coherences associated to each transition. The real and imaginary parts of these coherences are respectively related to the dispersion and absorption of the fields by the atomic media.

\subsection{Covariance Matrix dynamics and Spectral density}
The next step is the evaluation of the atomic covariances, since they are directly related with the  intensity fluctuations of the fields.
The evolution of the atomic covariance matrix $\mathbf{g}(t,t+\tau)=\langle \mathbf{x}(t),\mathbf{x}^\dagger(t+\tau) \rangle=\langle \mathbf{x}(t)\mathbf{x}^\dagger(t+\tau) \rangle - \langle \mathbf{x}(t) \rangle \langle \mathbf{x}^\dagger(t+\tau) \rangle$ can be obtained with the help of the dynamical equation (\ref{langevin1}) for  $\mathbf{x}(t)$. 

Let us start by describing the dynamical evolution of the covariance matrix for $\tau \neq 0$ based on the regression theorem ~\cite{Gardiner90}, such that 
\begin{subequations}
\begin{align}
\frac{d\mathbf{g}(\tau)}{d\tau}&=-\mathbf{M}\ \mathbf{g}(\tau), \hspace{0.5cm}\tau>0\\
\frac{d\mathbf{g}(\tau)}{d\tau}&= \mathbf{g}(\tau)\  \mathbf{M}^\dagger \hspace{0.8cm}\tau<0,
\end{align} \label{dgdtau}
\end{subequations}
\noindent where we define $\mathbf{g}(\tau)\equiv \mathbf{g}(t,t+\tau)$, simplifying the notation. 
The solution is directly obtained from the stationary solution
\begin{align}
\mathbf{g}(\tau)_{ss}=\begin{cases}  e^{-\mathbf{M} \ \tau}\mathbf{g}(0)_{ss}\hspace{1cm}&\tau>0\\
\mathbf{g}(0)_{ss} e^{\mathbf{M}^\dagger \ \tau}\hspace{1cm}&\tau<0.\end{cases}
\label{gcovfinal2}
\end{align}
where the subindex $ss$ denotes the stationary solution for $\mathbf{g}(\tau)$. In such a regime, for any interval of 
time $\tau$, the covariance matrix $\mathbf{g}(\tau)_{ss}$  is invariant under temporal displacements
i.e $\mathbf{g}(t,t+\tau)_{ss}=\mathbf{g}(t',t'+\tau)_{ss}$ for $t'\neq t$. The solution (\ref{gcovfinal2}) 
shows that, in stationary regime, the covariances between the atomic density elements present an exponential
decrease for intervals of time separated by $\tau$.

Now, let us consider the case when $\tau =0$. The dynamics is obtained by direct derivation of $\mathbf{g}(t,t)$ and application of Ito's calculus, resulting in  
\begin{align}
\frac{d\mathbf{g}(t,t)}{dt}&=-[\mathbf{M}\ \mathbf{g}(t,t) + \mathbf{g}(t,t)\  \mathbf{M}^\dagger] \nonumber\\
&+ 2\bar{\gamma} \Phi\ \mathbf{g}(t,t)\ \Phi + 2\bar{\gamma} \Phi \langle\mathbf{x}(t)\rangle \langle\mathbf{x}(t)^\dagger\rangle \Phi
\label{dgdt}
\end{align}
with 
\begin{equation}
\Phi=(\mathbf{N}_1+\mathbf{N}_2).
\label{Phi}\end{equation}

 In the stationary regime, we have
\begin{align}
2\bar{\gamma} \,\, \Phi \langle\mathbf{x}(t)\rangle_{ss} \langle\mathbf{x}(t)^\dagger\rangle_{ss} \Phi=&[\mathbf{M}\ \mathbf{g}(0)_{ss} + \mathbf{g}(0)_{ss}\   \mathbf{M}^\dagger]\nonumber\\ -&2\bar{\gamma}\, \, \Phi\ \mathbf{g}(0)_{ss}\  \Phi.\label{g_ss}
\end{align}

Typical measurements are performed rather in the frequency domain than in the time domain, allowing to selectively avoid the contribution of noise from other sources in the spectroscopy.
In the frequency domain, the intensity fluctuations are described by the spectral density, obtained from a Fourier transform of the covariance matrix in eq.(\ref{gcovfinal2}), 
resulting in the spectral density matrix
\begin{align}
\mathbf{S}(\omega)=&\frac{1}{2\pi}\left[(i\omega  + \mathbf{M})^{-1}\ \mathbf{g}(0)_{ss}\  + \mathbf{g}(0)_{ss}\ (-i\omega+\mathbf{M}^\dagger)^{-1}\right].
\label{Smatrix}
\end{align}
Computing $[i\omega  + \mathbf{M}]\mathbf{S}(\omega)[-i\omega  + \mathbf{M}^\dagger]$ and with the help of eq. (\ref{g_ss}), we obtain
\begin{align}
\mathbf{S}(\omega) =&\frac{1}{\pi}\left[(\mathbf{M}+i\omega)^{-1}\  \Phi\  \Pi\  \  \Phi\ (\mathbf{M}^\dagger - i\omega)^{-1}\right], \label{Smatrix2}
\end{align}
where
\begin{align}
\Pi =&\bar{\gamma}\ [\langle\mathbf{x}(t)\rangle_{ss} \langle\mathbf{x}(t)^\dagger\rangle_{ss}+ \mathbf{g}(0)_{ss}]. \label{Pimatrix}
\end{align}

We can relate the Fourier transform of the atomic covariance matrix described above (\ref{Smatrix}) to the intensity spectrum of the light fields given in eq. (\ref{SpectrInt}) 
using the thin sample model for the transmitted intensity relating fields and atomic coherences weighted by the effective optical density $\varrho_i$. 
Therefore the spectral correlation $C(\omega)$ in eq.~(\ref{eq:corfreq}) is given by spectral densities
 \begin{subequations}
\begin{align}
 S^I_{11}(\omega) &=\mu_{11} \{[\mathbf{S}(\omega)]_{33}+[\mathbf{S}(\omega)]_{44} -[\mathbf{S}(\omega)]_{34}-[\mathbf{S}(\omega)]_{43}\}, \label{S11}\\
 S^I_{22}(\omega) &=\mu_{22} \{[\mathbf{S}(\omega)]_{55}+[\mathbf{S}(\omega)]_{66} -[\mathbf{S}(\omega)]_{56}-[\mathbf{S}(\omega)]_{65}\},\label{S22}\\
 S^I_{12}(\omega) &=\frac{\mu_{12}}{2} \left.\{[\mathbf{S}(\omega)]_{36}+[\mathbf{S}(\omega)]_{45} -[\mathbf{S}(\omega)]_{35}-[\mathbf{S}(\omega)]_{46}\right.\nonumber \\
 &\left. +[\mathbf{S}(\omega)]_{63}-[\mathbf{S}(\omega)]_{64}+[\mathbf{S}(\omega)]_{54}-[\mathbf{S}(\omega)]_{53}\right\},\label{S12}
\end{align}\label{Sij}
\end{subequations}
where $\mu_{ij}=\varrho_i \mathcal{E}_i \mathcal{E}_j$. 

We can see that intensity spectra become a powerful tool to understand the atomic covariance matrix which is directly dependent on the coherences and populations involved in the EIT process.
The spectral density in eq.(\ref{Smatrix2}) takes the same form of an Ornstein-Uhlenbeck (OU) process, where $\Phi\ \Pi\ \Phi $ represents the diffusion matrix.
The matrix $\Pi$ (eq.~\ref{Pimatrix}) presents a contribution from the mean value of the atomic population and coherences $\langle \mathbf{x}_{ss}(t)\rangle$  (associated to the absorption and dispersion of the fields)
and their respective covariances $\mathbf{g}(0)_{ss}$. 

The resulting calculation, although successful in describing the observed spectra  presented in~\cite{Cruz07,HMFlorez13}, does not distinguish the contribution of each term in matrix $\Pi$.
So far, the calculation to obtain the spectral density matrix $\mathbf{S}(\omega)$, and consequently the intensity correlation $C(\omega)$, is reduced to a numerical task lacking analytical
and physical insight of the atomic response in the induction of a certain level of noise and intensity correlation. The perturbative approach we present below brings new information about the role of each term.

\section{Perturbative approach}\label{sec:PertModel}

The time evolution of the atomic variables in eq.(\ref{langevin1}) shows the relation between the optical pumping (included in the $\mathbf{a}$ matrix and proportional to the spontaneous decay $\Gamma$) and the laser linewidth (associated to $d\phi$) that determines how the system is affected 
by the excess of phase noise  in the incident light fields.
We may associate a perturbative parameter  to the laser linewidth as $\epsilon =\sqrt{\bar{\gamma}}$
and redefine the phase fluctuation as $d\phi_i(t)=\epsilon dW_i(t)$,
where  $dW_i(t)$ is a Wiener processes with $\langle dW_i(t),dW_j(t)\rangle=1$.
Then, following the perturbative approach described in \cite{Gardiner90}, we expand the vector $\mathbf{x}(t)$  as
\begin{align}
 \mathbf{x}(t) = \mathbf{x}^{(0)}(t) + \epsilon\ \mathbf{x}^{(1)}(t) + \epsilon^2 \mathbf{x}^{(2)}(t) + \cdots\ \label{ExpanPert}
\end{align}
After substitution of this result in eq.(\ref{x_ss}) we obtain the stationary solution for the mean value of the different orders $\mathbf{x}^{(n)}(t)$:
\begin{subequations}
\begin{align}
\langle\mathbf{x}^{(0)}(t)\rangle_{ss} &=\mathbf{M}^{-1}\ \mathbf{x}_o \label{x0_ss},\\
\langle\mathbf{x}^{(n)}(t)\rangle_{ss} &=0 \hspace{1cm} n\geq 1,
\end{align}
\end{subequations}
retaining nonzero solutions of the mean value only for the contribution of lowest order.

As for the covariance matrix $\mathbf{g}(t,t)$, it can be expanded in powers of $\epsilon$ as 
\begin{align}
\mathbf{g}(0)= \epsilon^2 \sigma^{(2)}+ \epsilon^3 \sigma^{(3)} + \epsilon^4 \sigma^{(4)} +  \epsilon^5 \sigma^{(5)}+\cdots\ ,
 \label{gcov}
\end{align}
where we define the covariances $\sigma^{(n)}$ as
\begin{subequations}
\begin{align}
\sigma^{(2)}=&\langle \mathbf{x}^{(1)}(t),\mathbf{x}^{(1)}(t)^\dagger \rangle ,\\
\sigma^{(3)}=&\langle \mathbf{x}^{(1)}(t),\mathbf{x}^{(2)}(t)^\dagger \rangle+\langle \mathbf{x}^{(2)}(t),\mathbf{x}^{(1)}(t)^\dagger \rangle ,\\
\sigma^{(4)}=&\langle \mathbf{x}^{(2)}(t),\mathbf{x}^{(2)}(t)^\dagger \rangle+ \langle \mathbf{x}^{(3)}(t),\mathbf{x}^{(1)}(t)^\dagger \rangle +\text{c.c},\\
\sigma^{(5)}=&\langle \mathbf{x}^{(2)}(t),\mathbf{x}^{(3)}(t)^\dagger \rangle+\langle \mathbf{x}^{(4)}(t),\mathbf{x}^{(1)}(t)^\dagger \rangle+\text{c.c},\\
\vdots \hspace{.5cm} & \hspace{1cm}  \vdots \hspace{3cm} \vdots\qquad . \nonumber
\end{align}
\end{subequations}
Substituting these results in eq.(\ref{g_ss}), we derive a recursive  formula to obtain the stationary condition for $\sigma^{(n)}$ matrices
\begin{subequations}
\begin{align}
\mathbf{M}\sigma_{ss}^{(2n)} + \sigma_{ss}^{(2n)}\mathbf{M}^\dagger=&2\  \Phi \ \sigma_{ss}^{2(n-1)}\ \Phi \hspace{0.6cm}n\geq1, \label{sigma_n}\\
\sigma_{ss}^{(2n-1)}=&0, \hspace{2.3cm} n\geq 2.  
\end{align}
\end{subequations}

All odd-order matrices vanish, while even-order matrices are recursively obtained from the zero-order matrix, given by the product of the vectors 
\begin{align}
\sigma_{ss}^{(0)}=\langle \mathbf{x}_{ss}^{(0)}\rangle\langle {\mathbf{x}_{ss}^{(0)}}^\dagger \rangle, \label{sigma_0}
\end{align}
which is calculated from the stationary solution given in eq.~(\ref{x0_ss}).

From the combined expansion of $\mathbf{g}(0)$ and the evaluation of the mean value  $\langle \mathbf{x}_{ss}^{(0)}\rangle$, eq. (\ref{Pimatrix}) can now be written as
\begin{align}
\Pi =&\sum_{n=0}^\infty \bar{\gamma}^{n+1} \sigma_{ss}^{(2n)},\label{Pi_expand}
\end{align}
involving a power series on the laser linewidth $\bar{\gamma}$.

Regarding the spectral density matrix $\mathbf{S}(\omega)$, from the substitution of 
 eq.(\ref{Pi_expand}) into eq.(\ref{Smatrix2}), we can also obtain an expansion involving increasing powers of the linewidth
\begin{align}
[\mathbf{S}(\omega)]_{kl}= \sum_{n=0}^\infty \bar{\gamma}^{n+1} [\mathbf{S}^{(2n)}(\omega)]_{kl} \label{SijExpand}
\end{align}
where
\begin{align}
\mathbf{S}^{(2n)}(\omega)=&\frac{1}{\pi}\left[(\mathbf{M}+i\omega)^{-1}\  \Phi\  \sigma_{ss}^{(2n)}\  \Phi^\dagger (\mathbf{M}^\dagger - i\omega)^{-1}\right]
\end{align}
This solution is one of the main results in this work. The matrix $\mathbf{S}^{(0)}(\omega)$ represents the leading term and contains the absorptive and dispersive properties of the medium from eq.(\ref{sigma_0}). Higher order terms are considered for a complete description of the system.

\subsection{Correlation spectroscopy}

\begin{figure}[htb!]
\centering
\includegraphics[width=8.6cm]{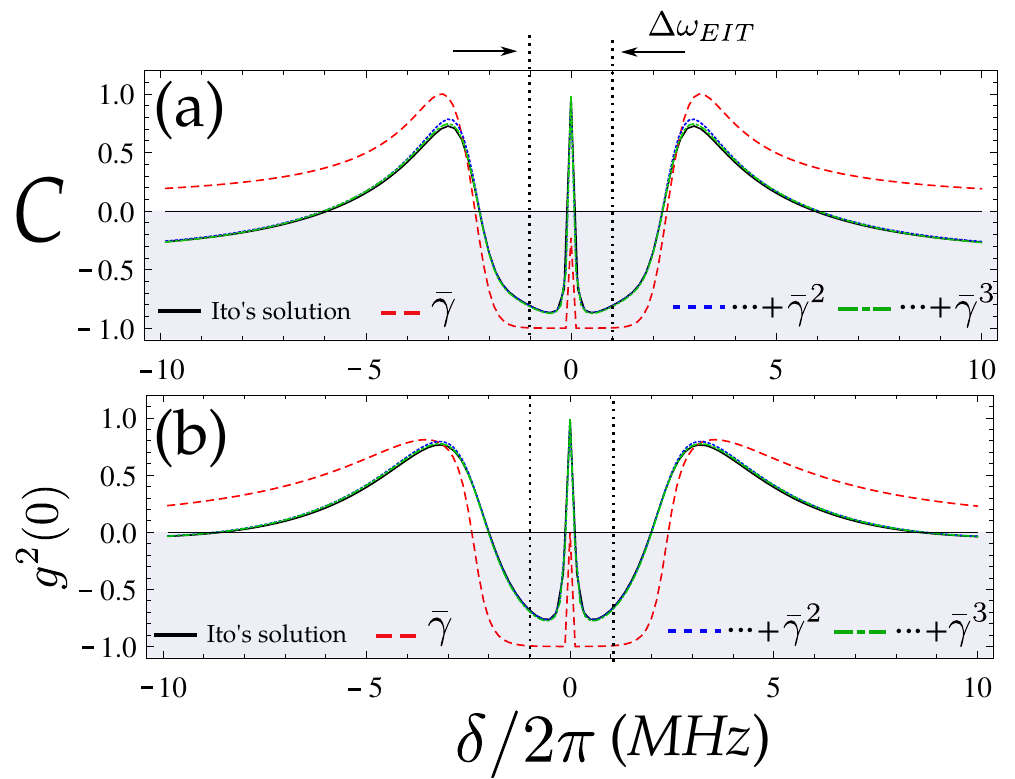}
\vspace{-.8cm}
\caption{(Color online) Intensity correlation as a function of the two-photon detuning $\delta=\Delta_1-\Delta_2$ (see Fig.\ref{fig:Levels}(a)) of two beams interacting with a three-level $\Lambda$ system.
(a) Correlation coefficient $C(\omega)$ and (b) correlation function $g^2(0)$. The transparency window is $\Delta\omega_{EIT}/2\pi\sim 2$MHz. Parameters for the calculation: laser linewidth $\bar{\gamma}/2\pi=1$MHz, Rabi frequency $\Omega_1=\Omega_2=0.3\Gamma$, 
ground states decoherence $\gamma_d/2\pi=150$kHz, natural linewidth $\Gamma/2\pi= 6$MHz, analysis frequency $\omega/2\pi=2$MHz and one-photon detuning $\Delta_2=0$. }
\label{fig:CPert}
\end{figure}

The perturbative expansion of the noise spectra and the cross-correlation term in eq.(\ref{Sij}) are obtained from substitution of the matrix elements from eq.(\ref{SijExpand}).
Therefore the intensity correlation $C(\omega)$ in eq.(\ref{eq:corfreq}) has different contributions depending on the terms we are taking into account in the expansion of the spectra.
The same applies to the evaluation of $g^2(0)$ (eq.~\ref{eq:cor}) in the perturbative approach. Figs.~\ref{fig:CPert}(a) and (b) show the perturbative result for correlation spectra $C(\omega)$  and $g^2(0)$, 
respectively, considering  the same anti-correlation regime studied in refs.\cite{Felinto13,HMFlorez13}, with laser linewidth of $\bar{\gamma}/2\pi=1$MHz. 
The dashed line represents the first order expansion in $\bar{\gamma}$ associated to $\sigma^{(0)}$. Corrections of  higher order terms  up to $\bar{\gamma}^2$ and $\bar{\gamma}^3$ are shown
in dotted and dotted-dashed line, respectively. The convergence to Ito's solution (solid line, as in ref. \cite{HMFlorez13}) is evident. It is clear that considering the expansion up to 
$\bar{\gamma}^2$ would be a valid approximation to describe the intensity correlation in this case. 

\begin{figure}[htb!]
\centering
\includegraphics[width=8.6cm]{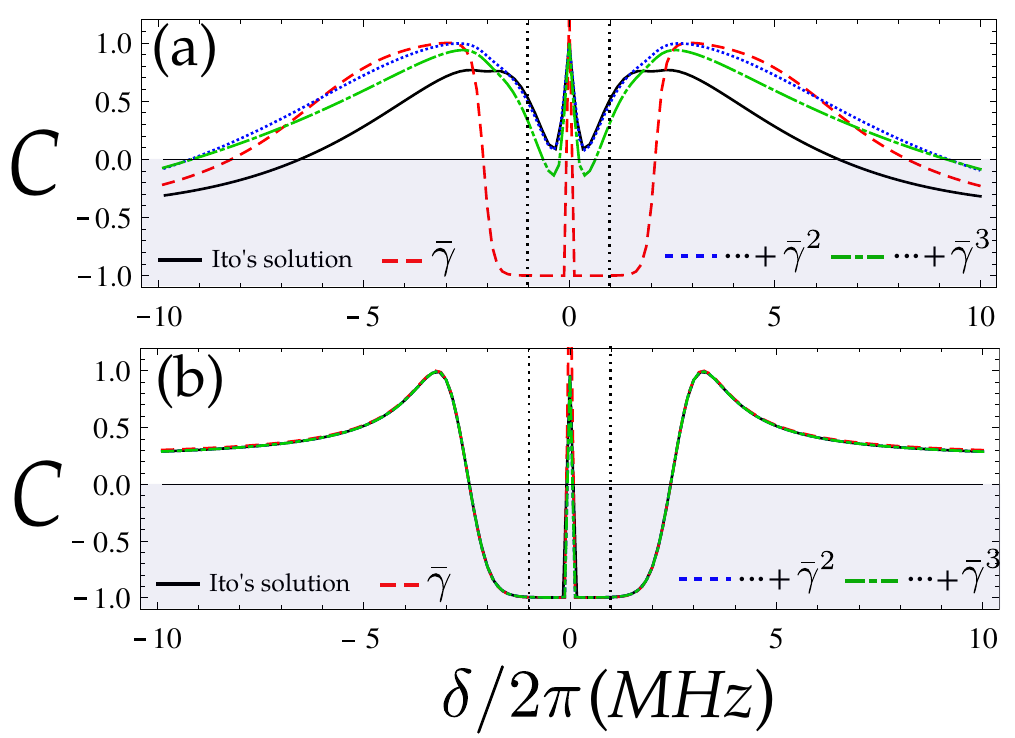}
\vspace{-.5cm}
\caption{(Color online)  Intensity correlation spectra for a  laser linewidth: (a) $\bar{\gamma}/2\pi=10$MHz  and (b) $\bar{\gamma}/2\pi=0.01$MHz. The parameters for the calculation are the same as those in Fig. \ref{fig:CPert}. The two vertical dotted lines represent the transparency window $\Delta\omega_{EIT}$.}
\label{fig:ClowPert}
\end{figure}

The limit of relevant terms in the evaluation of the correlation is obviously dependent on the laser linewidth, as is shown in 
Figs.~\ref{fig:ClowPert}(a) and (b). The correlation spectra for laser linewidths of the order of $\bar{\gamma}/2\pi=10$MHz (high noise laser) 
and $\bar{\gamma}/2\pi=0.01$MHz (low noise laser) are compared. For a high noise laser, Fig.~\ref{fig:ClowPert}(a) shows that higher 
order terms are necessary to determine completely the intensity correlation.

 On the other hand, Fig.~\ref{fig:ClowPert}(b) shows that for low noise lasers, 
the intensity correlation is fully described by the first order term, which matches Ito's solution. 
This demonstrates  that for a coherent light source, the usual linearization approximation for the atomic variables 
$\mathbf{x}(t)=\langle\mathbf{x}(t)\rangle + \delta \mathbf{x}(t)$  is valid and sufficient.

In both cases, the intrinsic linewidth of the correlation spectrum remains narrower than the transparency window 
$\Delta\omega_{EIT}$ measured by the FWHM of the absorption $\text{Im} \langle \rho_{i3}\rangle_{ss}$. It was shown in refs. 
\cite{Xiao09,Felinto13,HMFlorez13} that, unlike the transparency window $\Delta \omega_{EIT}$ which is sensitive to power
broadening, the linewidth of the correlation spectrum is free from power broadening and is proportional to the ground
states' decoherence rate. What we can observe in Figs.~\ref{fig:ClowPert}(a) and (b) is that the intrinsic linewidth
for lower and higher order terms is always narrower than $\Delta \omega_{EIT}$, because it is mainly dependent on the
ground states decoherence.

These results encourage the search for an explicit expression for the correlation $C(\omega)$, in the same way that eq.~(\ref{g20}) 
was evaluated in ref.~\cite{Felinto13}. Our goal is to find an expression for the intensity correlation written in terms of the
absorption and dispersion of the light fields, in the frequency domain. 

\section{Mapping the absorption and dispersion to the correlation spectra}\label{sec:Mapping}

The perturbative expansion of the $\Pi$ matrix in eq.(\ref{Pi_expand}), together with the recursive evaluation of covariance 
matrices given by eq.~(\ref{sigma_n}), shows how the first order term, which depends directly on the absorptive and dispersive 
response of the atoms, establishes the leading term for the spectral density. From that first response we can determine higher order corrections. 
Hence, by expressing the $\Pi$ matrix in terms of the real and imaginary part of the coherences,  we can decompose the spectral density
explicitly in terms of absorption and dispersion, as in the heuristic model in eq.(\ref{g20}).

The matrix $\sigma^{(0)}$ defined in eq.(\ref{sigma_0}) is calculated  from the zero order term $\langle \mathbf{x}_{st}^{(0)}\rangle$ 
that has its  stationary solution given in eq.(\ref{x0_ss}). It is convenient to shift to a new vector 
$\tilde{\mathbf{x}}=(\rho_{11}^{(0)},\rho_{22}^{(0)},\text{Re} p_1,\text{Im} p_1,\text{Re} p_2$, $\text{Im} p_2,\rho_{12}^{(0)},\rho_{21}^{(0)})$
explicitly dependent on the absorption and dispersion of the fields. This can be done by the transformation
\begin{align}
\tilde{\mathbf{x}}=\mathbf{U} \langle \mathbf{x}_{st}^0\rangle=
\begin{bmatrix}
1&0&0&0&0&0&0&0\\
0&1&0&0&0&0&0&0\\
0&0&\frac{1}{2}& \frac{1}{2}&0&0&0&0\\
0&0&-\frac{i}{2}& \frac{i}{2}&0&0&0&0\\
0&0&0&0&\frac{1}{2}&\frac{1}{2}&0&0\\
0&0&0&0&-\frac{i}{2}&\frac{i}{2}&0&0\\
0&0&0&0&0&0&1&0&\\
0&0&0&0&0&0&0&1&\\
\end{bmatrix}\begin{bmatrix}
\rho_{11}^{(0)}\\
\rho_{22}^{(0)}\\
p_1\\
p_1^*\\
p_2\\
p_2^*\\
\rho_{12}^{(0)}\\
\rho_{21}^{(0)}\\
\end{bmatrix}.\label{TransReIm}
\end{align}

This transformation is helpful to obtain the noise spectra for each beam in terms of 
the absorption and dispersion associated to each transition. 
The spectral density matrix given by eq.~(\ref{Smatrix2}) can be evaluated from the expansion of covariances described in  eq.~(\ref{SijExpand}).
We begin by the evaluation of the covariances in the rotating frame, using eq.~(\ref{Phi})
\begin{align}
\Phi\   \sigma^{(n)} \Phi=
\begin{bmatrix}
0&0&0&0&0&0&0&0\\
0&0&0&0&0&0&0&0\\
0&0&\sigma_{33}^{(n)}& -\sigma_{34}^{(n)}&\sigma_{35}^{(n)}&-\sigma_{36}^{(n)}&0&0\\
0&0&-\sigma_{43}^{(n)}& \sigma_{44}^{(n)}&-\sigma_{45}^{(n)}&\sigma_{46}^{(n)}&0&0\\
0&0&\sigma_{53}^{(n)}& -\sigma_{54}^{(n)}&\sigma_{55}^{(n)}&-\sigma_{56}^{(n)}&0&0\\
0&0&-\sigma_{63}^{(n)}& \sigma_{64}^{(n)}&-\sigma_{65}^{(n)}&\sigma_{66}^{(n)}&0&0\\
0&0&0&0&0&0&0&0&\\
0&0&0&0&0&0&0&0&\\
\end{bmatrix}\label{sigma_nPert}
\end{align}
resulting in a reduced number of relevant independent elements.

Now using the transformation of eq.(\ref{TransReIm}) into the $\Pi$ matrix, its expansion becomes 
\begin{align}
\tilde{\Pi} &=\mathbf{U}  \Phi\  \Pi\  \Phi\ \mathbf{U}^{-1}, \nonumber\\
&=\sum_{n=0}^\infty \bar{\gamma}^{n+1} \tilde{\sigma}^{(2n)},\label{Pi_transfm1}
\end{align}
where every  $\tilde{\Pi}_{ij}$ matrix element is explicitly written as the perturbative corrections performed  around
the mean value  $\tilde{\sigma}^{(0)}_{ij}$, as described in the appendix \ref{apendA}, in eq.(\ref{CovHeuristica_trans}).
Each term $\tilde{\sigma}^{(n)}=\mathbf{U}  \Phi\  \sigma^{(n)}\  \  \Phi\ \mathbf{U}^{-1}$ takes the same  form as the matrix (\ref{sigma_nPert}). 

 The noise spectra in eqs.(\ref{S11})-(\ref{S12}) are then given by products of the expansion of the covariances $\tilde\Pi_{ij}$ by weighting factors that depend on the analysis frequency~$\omega$:
\vspace{-.2cm}\begin{align}
S^I(\omega)_{11}& = \sum_{i,j=3}^6 \alpha_{ij}(\omega)\ \tilde{\Pi}_{ij}\label{s11_pert_decomp}\\
S^I(\omega)_{22}& = \sum_{i,j=3}^6 \beta_{ij}(\omega)\ \tilde{\Pi}_{ij}\label{s22_pert_decomp}\\
S^I(\omega)_{12}& = \sum_{i,j=3}^6 \nu_{ij}(\omega) \tilde{\Pi}_{ij}\label{s12_pert_decomp}
\end{align}
A detailed calculation of the weighting coefficients is found in appendix \ref{apendA}, where we present the detailed 
 expression for the noise power spectra for each field and their correlation.

If we keep only the expansion to lowest order, an explicit expression for the normalized spectral  correlation is 
\begin{widetext}
\begin{equation}
C(\omega) =\frac{\tilde{\nu}_{35}(\omega)\text{Im} p_1\text{Im}p_2 +\tilde{\nu}_{46}(\omega)\text{Re}p_1\text{Re}p_2 
+\tilde{\nu}_{36}(\omega) \text{Im}p_1\text{Re}p_2+\tilde{\nu}_{45}(\omega)\text{Im}p_2\text{Re}p_1
 +C_{12}(\omega) }
{\sqrt{\left[ \alpha_{33}(\omega)\text{Im}^2 p_1+\alpha_{44}(\omega)\text{Re}^2p_1
+C_{11}(\omega)\right] \left[ \beta_{55}(\omega)\text{Im}^2p_2 +\beta_{66}(\omega) \text{Re}^2p_2+C_{22}(\omega) \right] }}. \label{CPertDecom}
\end{equation}
\end{widetext}

This result is similar to the correlation function from the heuristic model  in eq.(\ref{g20}). But instead of having  
$\text{Im}\  p_1 \text{Im}\  p_2$ and $\text{Re}\  p_2 \text{Re}\  p_1$ with equal contributions to the correlation, a 
weighting factor $\nu_{35}(\omega)$ and $\nu_{46}(\omega)$ modulates their contribution in eq.(\ref{CPertDecom}).  

Moreover, beyond the contribution from  $\text{Im}\ p_1 \text{Re}\ p_2$ and $\text{Im}\ p_2 \text{Re}\ p_1$, all possible 
cross terms from matrix ~(\ref{CovHeuristica_trans}) will be present in evaluation of $S^I_{12}$, $S^I_{11}$ and $S^I_{22}$. 
The contribution of the terms not explicitly shown in eq.~(\ref{CPertDecom}) is given by the weaker but relevant contributions
$C_{11}$, $C_{22}$ and $C_{12}$ defined in eqs.~(\ref{C_11})-(\ref{C_12}).
 
\begin{figure}[htb!]
\centering
\includegraphics[width=8.6cm]{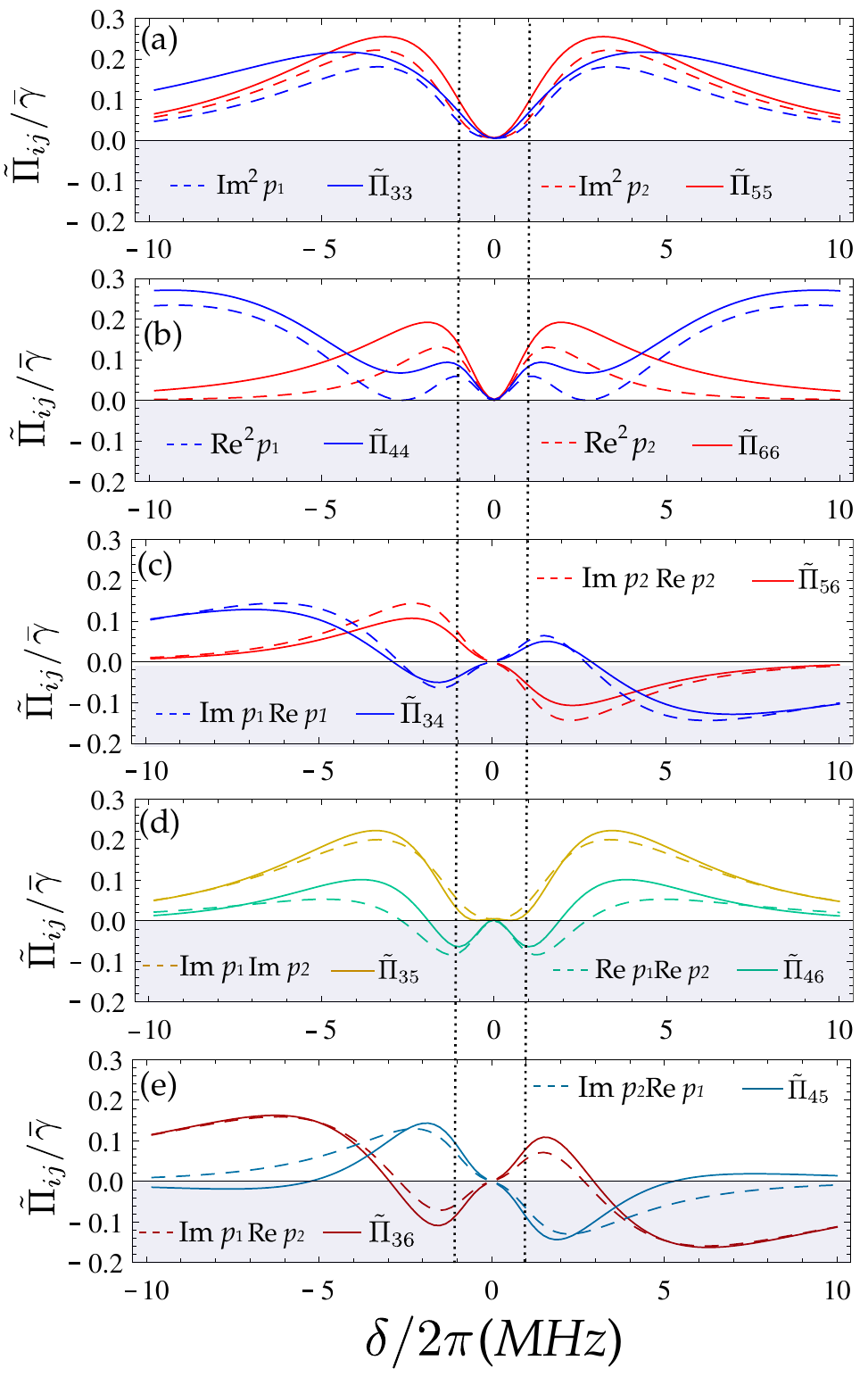}
\vspace{-.5cm}
\caption{(Color online) Comparing the $\tilde{\Pi}_{ij}$ matrix elements at first order ($\sigma_{ij}^{(0)}$) with higher order terms in the perturbative expansion   $\tilde{\Pi}_{ij}/\bar{\gamma}  =\sigma_{ij}^{(0)}+\cdots+\bar{\gamma}^{6}\sigma_{ij}^{(7)}$. (a) Absorptive and  (b) dispersive response for each beam. (c) Absorption $\times$ Dispersion product  for each beam. (d) and (e) cross terms products $\text{Im}p_1\text{Im}p_2$ ($\text{Re}p_1\text{Re}p_2$) and $\text{Im}p_1\text{Re}p_2$ ($\text{Re}p_1\text{Im}p_2$). The parameters for the calculation are the same as in Fig. \ref{fig:CPert}. The two vertical dotted lines represent $\Delta\omega_{EIT}$.}
\label{fig:PiCorrections}
\end{figure}

In Fig.~\ref{fig:PiCorrections} we observe the effect of the perturbative corrections taken into account in the evaluation
of the normalized $\tilde{\Pi}_{ij}$ matrix elements, considering typical values of $\bar{\gamma}/2\pi=1$MHz and $\Delta_2=0$. 
Figures (a), (b) and (c) show the effect of perturbative corrections $\text{Im}^2p_i$, $\text{Re}^2p_i$ and $\text{Im}p_i\text{Re}p_i$, 
respectively, for the two beams $i=1,2$. Figures (c) and (d) show the perturbative corrections for 
$\text{Im}p_1\text{Im}p_2$ ($\text{Re}p_1\text{Re}p_2$) and for the products $\text{Im}p_1\text{Re}p_2$ ($\text{Re}p_1\text{Im}p_2$). 
We may observe that the lowest order terms closely follow $\tilde{\Pi}_{ij}/\bar{\gamma}$. 

Now, comparing  the dispersive and the absorptive response, Figs.~\ref{fig:PiCorrections}(a) and (b), show that 
$\text{Im}^2p_i\sim \text{Re}^2p_i$ for $|\delta|<\Delta\omega_{EIT}$. Now, for $|\delta|> \Delta\omega_{EIT}$, 
the atomic response presents situations where the absorption overcomes the dispersion, or vice versa. 
The product $\text{Im}\ p_i \text{Re}\ p_i$ also plays a role in the atomic response for each field, following a dispersive profile in 
(c). The coefficients $\alpha_{ij}(\omega)$, $\beta_{ij}(\omega)$ and $\nu_{ij}(\omega)$ will determine the leading term in each specific situation.

On the other hand, the cross-product $\text{Re}p_1\text{Re}p_2$ competes with $\text{Im}p_1\text{Im}p_2$ for 
$|\delta|<\Delta\omega_{EIT}$ as shown in Fig.~\ref{fig:PiCorrections}(d). Only the coefficients $\nu_{ij}(\omega)$ will determine the main contribution for the induced correlation or anti-correlation. 
In the case of $|\delta|/2\pi> 6$MHz, the absorptive ($\tilde{\Pi}_{35}$) and dispersive ($\tilde{\Pi}_{46}$) terms are positive and contribute to $C(\omega)>0$. 

Finally, Fig.~\ref{fig:PiCorrections}(e) shows the cross-products $\text{Im}p_i\text{Re}p_j$ that present a dispersion-like profile. 
Such terms are absent in the evaluation of $g^2(0)$. 
Its interesting to notice that the contribution of higher order terms is important in fliping the sign of   $\tilde{\Pi}_{45}$ 
 for $|\delta|/2\pi> 6$MHz. 
  This term is 
responsible for inverting the sign of the spectral correlation in Fig.~\ref{fig:CPert}(a) for higher detuning, when higher
order terms are taken into account.

\begin{figure}[htb!]
\centering
\includegraphics[width=8.6cm]{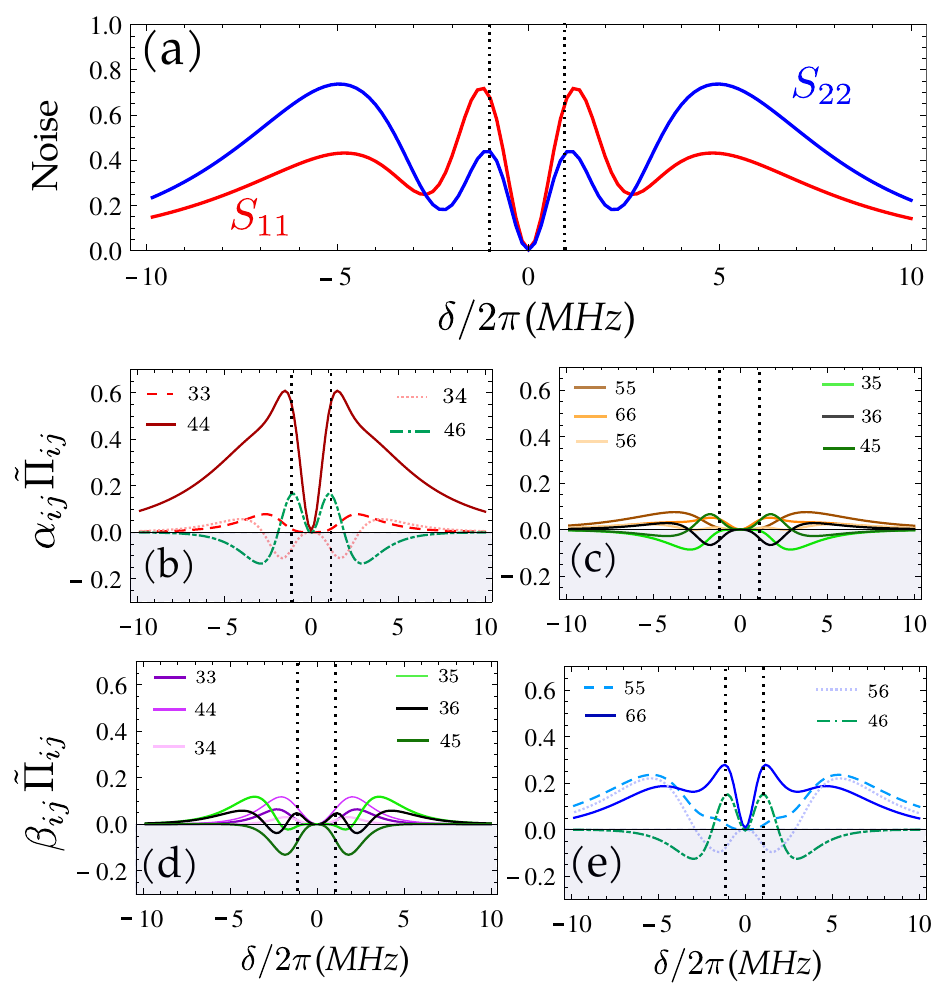}
\vspace{-.5cm}
\caption{(Color online) Decomposing the noise spectra $S_{11}$ and $S_{22}$  in terms of the $\alpha_{ij}$ and $\beta_{ij}$ coefficients, 
respectively. The noise spectra in (a) are calculated up to order $\bar{\gamma}^{7}$ for analysis frequency $\omega/2\pi=2$ MHz. 
(b) and (e) show the main coefficients $\alpha_{ij}$ and $\beta_{ij}$ respectively. (c)  shows the weaker coefficients
$\alpha_{55}$, $\alpha_{66}$, $\alpha_{56}$, $\alpha_{35}$, $\alpha_{36}$ and $\alpha_{45}$
  (d)  shows the weaker coefficients $\beta_{33}$, $\beta_{44}$, $\beta_{34}$, $\beta_{35}$, $\beta_{36}$ and $\beta_{45}$. 
  The parameters for the calculation are the same that in Fig. \ref{fig:CPert}. The two vertical dotted lines represent $\Delta\omega_{EIT}$. }
\label{fig:Noise}
\end{figure}

The coefficients $\alpha_{ij}(\omega)$, $\beta_{ij}(\omega)$ and $\nu_{ij}(\omega)$ will determine the effective contribution of the
$\tilde{\Pi}_{ij}$ matrix elements to the noise spectra and their intensity correlation. Let us start by the decomposition of the noise 
of each light beam. In Fig.~\ref{fig:Noise}(a) we plot the noise spectrum $S_{11}(\omega)$  and $S_{22}(\omega)$ evaluated with terms 
up to order $\bar{\gamma}^{7}$ in $\tilde{\Pi}$. In Figs.~\ref{fig:Noise}(b) and (c) we plot the contribution of the product 
$\alpha_{ij} \Pi_{ij}$ associated to $S_{11}$ and the Figs.~\ref{fig:Noise}(d) and (e) show the products $\beta_{ij}\Pi_{ij}$ 
associated to $S_{22}$. 

Unlike the heuristic model of $g^2(0)$, where $\text{Im}^2p_i$ and $\text{Re}^2p_i$ contribute equally to the noise of the beams,
Figs.~\ref{fig:Noise}(b) and (e) show that the dispersive response ($\tilde{\Pi}_{44}$ and $\tilde{\Pi}_{66}$) overcomes the
absorptive one ($\tilde{\Pi}_{33}$ and $\tilde{\Pi}_{55}$), since $\alpha_{44}\tilde{\Pi}_{44}> \alpha_{33}\tilde{\Pi}_{33}$ and 
$\beta_{66}\tilde{\Pi}_{66}> \beta_{55}\tilde{\Pi}_{55}$, respectively, for $|\delta|<\Delta\omega_{EIT}$. In particular, for $|\delta|\sim \Delta\omega_{EIT}$
the components $\alpha_{44}\tilde{\Pi}_{44}$ and $\beta_{66}\tilde{\Pi}_{66}$ become the main contributions to $S_{11}(\omega)$ and $S_{22}(\omega)$ in Fig. \ref{fig:Noise}.(a) among all the terms.
Nevertheless, we can notice that other terms will play an important role for $|\delta|\geq \Delta\omega_{EIT}$, like the contributions related to  $\alpha_{46}(\beta_{46})$, associated
to $\text{Re}p_1 \text{Re}p_2$ (Fig.~\ref{fig:PiCorrections}c). Moreover, for such detuning range, the absorptive term $\beta_{55}\tilde{\Pi}_{55}$ and dispersive term $\beta_{66}\tilde{\Pi}_{66}$
present almost the same contribution to the noise spectrum $S_{22}(\omega)$, different from $S_{11}(\omega)$ where the dispersion given by $\alpha_{44}\tilde{\Pi}_{44}$
still overcomes the absorption $\alpha_{33}\tilde{\Pi}_{33}$.
Other important terms modulated by $\alpha_{34}$ and $\beta_{56}$, associated to  $\text{Im}p_i \text{Re}p_i$ for $i=1,2$,
also present a noticeable effect.  Figs.~\ref{fig:Noise}(c) and (d) show the remaining coefficients $\alpha_{ij}\tilde{\Pi}_{ij}$ and $\beta_{ij}\tilde{\Pi}_{ij}$ that define additional terms 
in eqs.(\ref{s11_pert_decomp}) and (\ref{s22_pert_decomp}). Although, individually they are  apparently relevant terms, their sum represents  effectively a weaker contribution with respect 
to the coefficients in figures (b) and (e).

\begin{figure}[htb!]
\centering
\includegraphics[width=8.6cm]{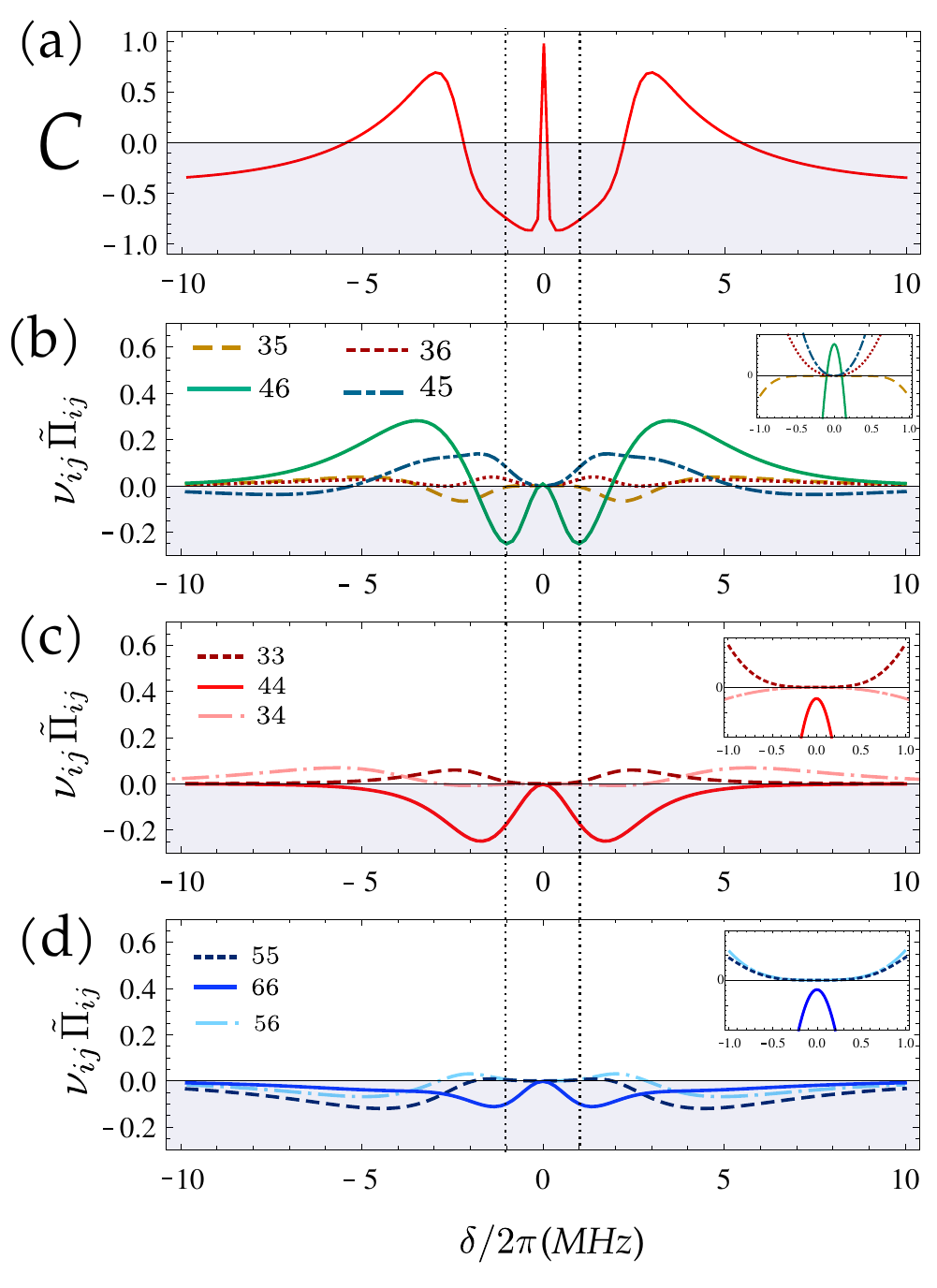}
\vspace{-.5cm}
\caption{(Color online) (a) Correlation coefficient. (b) and (c) Coefficients $\nu_{ij}(\omega)$ associated to the $\tilde{\Pi}_{ij}$ matrix elements to obtain the 
spectral density of the cross-correlation $S_{12}(\omega)$ for $\Delta_2=0$ and $\omega/2\pi=2$ MHz. The two vertical dotted lines represent $\Delta\omega_{EIT}$.}
\label{fig:Coeffabv}
\end{figure}

In order to get a better understanding of the profile of the spectral correlation $C(\omega)$, we analyze the intensity 
correlation given by $S_{12}$ (eq.~\ref{s12_pert_decomp}) associated to the $\nu_{ij}$ coefficients.
These coefficients weight the $\tilde{\Pi}_{ij}$ matrix elements used in evaluation of the intensity correlation shown in Fig.~\ref{fig:Coeffabv}.(a).
The effective contributions given by the products $\nu_{ij}\Pi_{ij}$ are presented in Fig.~\ref{fig:Coeffabv}.(b) and (c).
Unlike the heuristic model where the absorptive and  the dispersive response have the same weight to the final correlation,
Fig.~\ref{fig:Coeffabv}(b) shows that, in the frequency domain, the contribution from the dispersive part 
$\tilde{\nu}_{46}(\omega)[\text{Re}p_1\text{Re}p_2+\cdots$], is significantly greater than the  absorptive response 
$\tilde{\nu}_{35}(\omega)[\text{Im}p_1\text{Im}p_2+\cdots]$. 
It also shows that the contribution to the intensity correlation  $C(\omega)$ in eq.(\ref{CPertDecom}) depends strongly on 
$\tilde{\nu}_{45}(\omega)$ associated to the  product $\text{Im}p_2\text{Re}p_1$, plotted in Fig.~\ref{fig:PiCorrections}(d). 

Besides these differences between the $g^2(0)$ and $C(\omega)$, the spectral correlation also presents contribution from 
$\nu_{44}(\omega)[\text{Re}^2p_1+\cdots]$ and $\nu_{66}(\omega)[\text{Re}^2p_2+\cdots]$, 
as shown in Fig.~\ref{fig:Coeffabv}(c).  The coefficients $\nu_{33}(\omega)$, $\nu_{55}(\omega)$, $\nu_{34}(\omega)$ and $ \nu_{56}(\omega)$
present a weaker contribution with respect to other coefficients, and are associated to correction $C_{12}$ in eq.~(\ref{CPertDecom}).
The combined result leads to  anti-correlation between the two beams for $|\delta|/2\pi<3$MHz. 
However, for $\delta=0$ the only non-zero terms that effectively contribute to correlation between the beams are  
$\tilde{\Pi}_{46}$, $\tilde{\Pi}_{44}$ and  $\tilde{\Pi}_{66}$ (insets in figures (b) and (c)).  Now, 
for $|\delta|/2\pi\sim 3.5$MHz the dispersive term $\nu_{46}\tilde{\Pi}_{46}$ takes its maximum value becoming the main 
contribution for a positive correlation C. 

On the other hand,  for detuning $|\delta|>\Gamma=(2\pi) 6$MHz, the competition among the non-zero terms 
leads to anti-correlation between the light fields.  However, it is worth noting that the sign inversion of  
$\nu_{45}(\omega)[\text{Im}p_2\text{Re}p_1+ \cdots]$ near $|\delta|\sim 5$MHz in Fig.~\ref{fig:Coeffabv}(b) 
determines the transition from correlation to anticorrelation for $C$ in Fig.~\ref{fig:Coeffabv}.(a).
This particular term can be understood as light being absorbed in the beam $\mathbf{E}_2$ ($\text{Im}p_2$), which is on resonance ($\Delta_{2}=0$), and re-emitted in the frequency mode of the field $\mathbf{E}_1$, depending strongly on its dispersion associated to $\text{Re}p_1$. Therefore we have non-resonant Stokes transitions, contributing to  anti-correlation.




\section{Sidebands Resonances broadening the correlation spectroscopy}\label{sec:sidebands}

\begin{figure}[htb!]
\centering
\includegraphics[width=8.6cm]{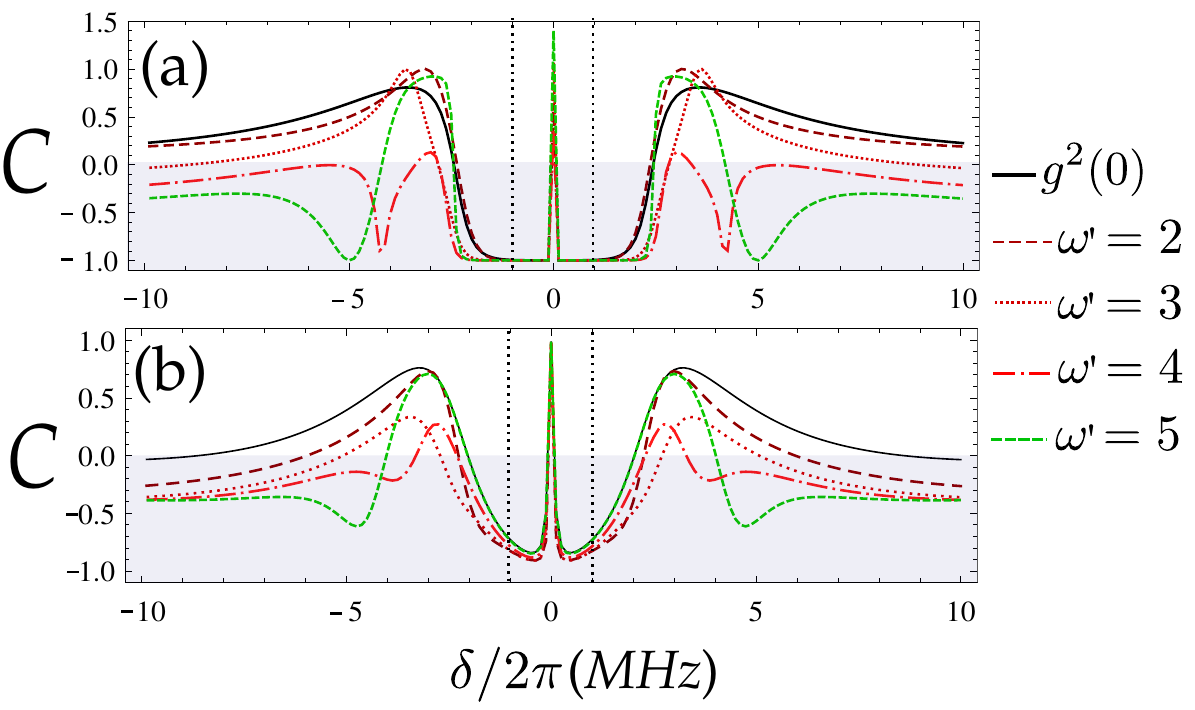}
\vspace{-.5cm}
\caption{(Color online)  Intensity correlation spectra for a  laser linewidth of  $\bar{\gamma}/2\pi=1$MHz at (a) First order $\bar{\gamma}$ and 
(b) considering the perturbative expansion up to $\bar{\gamma}^{7}$ in eq.(\ref{CPertDecom}).
It is considered $\omega'=\omega/2\pi=2$MHz, 3MHz, 4MHz and 5MHz. The parameters for the calculation are the same as those
in Fig. \ref{fig:CPert}. The two vertical dotted lines represent the transparency window $\Delta\omega_{EIT}$.}
\label{fig:CPert_wa}
\end{figure}

In this section we investigate the intensity correlation for different analysis frequencies. In Fig.~\ref{fig:CPert_wa} we compare the coefficient $C(\omega)$ and $g^2(0)$ function for $\omega/2\pi=2$MHz, 
3MHz, 4MHz and 5MHz, considering a standard noisy laser of $\bar{\gamma}/2\pi=1$MHz.  In Fig.~\ref{fig:CPert_wa}(a)  we plot the correlation coefficient with $\tilde\Pi$ evaluated to first order in $\bar{\gamma}$,
while in Fig.~\ref{fig:CPert_wa}(b) they are evaluated to the seventh order. Besides the unphysical result of $C>1$ on resonance $\delta=0$, the general profiles present reasonable similarities. 
One of them is the negative value for the correlation in the range of the EIT, except for the narrow structure, with the width of the inverse of the atomic coherence between the ground levels. 
The main difference is that the correlation is saturated in the linearized case, while a more careful treatment shows its limiting values.

Important features that we observe for higher analysis frequencies are the resolved resonances at $\delta=\pm\omega$ for  $\omega/2\pi=4$MHz
and 5MHz, while for smaller analysis frequencies, there is a broadening in the anticorrelation peak, when compared to the response of $g^2(0)$ on the two photon detuning.
On the other hand, for  $|\delta|<3\times\Delta\omega_{EIT} \sim (2\pi)\ 6$MHz, the correlation profile of $C(\omega)$ coincides with the $g^2(0)$ function (solid line) for
$\omega/2\pi=5$MHz. That is, we analyze the intensity  correlation at a particular frequency, such that sideband resonances do not influence
the PN-AN process inside $3\times\Delta\omega_{EIT}$. It is worth noting that for any analysis frequency considered, the intrinsic linewidth of 
$C(\omega)$ is not affected. This guarantees that the intrinsic linewidth measured in the frequency or the time domain is the same, 
no matter what analysis frequency is chosen \cite{Felinto13,HMFlorez13}. Finally, evaluation of higher order terms results 
in broadened (specially for $\omega/2\pi=4$MHz), and less anti-correlated resonances for the sidebands.

\begin{figure}[htb!]
\centering
\includegraphics[width=8.6cm]{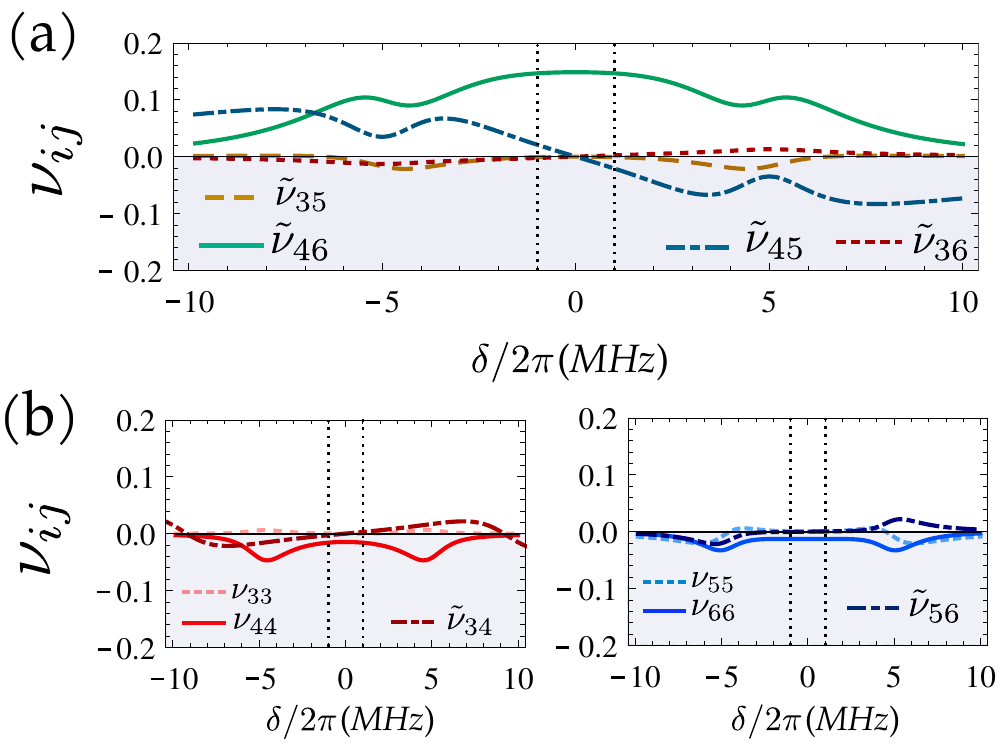}
\vspace{-.5cm}
\caption{(Color online) Coefficients $\nu_{ij}(\omega)$ that weigh the $\tilde{\Pi}_{ij}$ matrix elements to obtain the spectral density of the cross-correlation $S_{12}(\omega)$ for $\Delta_2=0$ and $\omega/2\pi=5$MHz. The two vertical dotted lines represent the transparency window $\Delta\omega_{EIT}$.}
\label{fig:Coeffabv_wa5MHz}
\end{figure}

In order to gain insight into the sideband resonances, we decompose the correlation in terms of the atomic response as in section \ref{sec:Mapping}. The $\nu_{ij}(\omega)$ 
coefficients are plotted in Fig.~\ref{fig:Coeffabv_wa5MHz} for $\omega/2\pi=5$MHz, where the sideband resonances are resolved.
As in the case of Fig.~\ref{fig:Coeffabv}, the main contributions for the correlation are given by the dispersive terms $\tilde{\Pi}_{46}$ and $\tilde{\Pi}_{45}$ for $\delta\neq\omega$. 
However, at the sideband resonances $\delta=\pm\omega$,  the contribution from both terms decreases, while the negative contribution from the $\tilde{\Pi}_{44}$ and $\tilde{\Pi}_{66}$ remains,
leading to anti-correlation. 

The decomposition in terms of the $\tilde{\Pi}_{ij}$ matrix elements and the coefficients $\nu_{ij}(\omega)$  
in Figs.~\ref{fig:PiCorrections} and ~\ref{fig:Coeffabv_wa5MHz}, respectively,  show the independent contribution of the 
PN-AN process when the carrier's two photon detuning is $|\delta|<\Delta\omega_{EIT}$ and $\omega/2\pi=4$MHz and 5MHz.
For $\omega/2\pi\leq 3$MHz the sideband resonances are not resolved, broadening the correlation profile in the frequency domain
for $|\delta|<\Delta\omega_{EIT}$.

 \section{conclusions}\label{sec:conclusions}

The mapping between the intensity correlation and the absorptive and dispersive properties of the atomic medium under EIT condition, 
was studied in the frequency domain. A perturbative model was proposed providing analytical elements to describe such a mapping. 
We obtained an expression for the correlation coefficient in terms of the absorption and dispersion of the light fields analogous  
to the heuristic model in ref.~\cite{Felinto13}. We have also shown the  explicit  dependence of the correlation coefficient on the
laser linewidth as it was observed in Ref.~\cite{Xiao14}. Since the perturbative parameter corresponds to the laser linewidth, 
such a dependence rises when higher order terms are take into account for determining  the correlation. It was shown that for low 
phase noise beams  the first order term describes completely their  spectral densities and the intensity correlation. 

Moreover we show that the intrinsic linewidth can be measured by the correlation spectroscopy, independently of the approach i.e.
in the time or frequency domain approach. We demonstrated that for analysis frequency of the order of the natural linewidth the 
correlation coefficient presents some particular resonances due to sidebands of the input light fields. In such condition the correlation
coefficient in eq.(\ref{CPertDecom}) can take the same form as the $g^2(0)$ function for detuning of the order of the transparency window.
 The physical insight given by the perturbative model can be experimentally studied by gradually applying phase noise into coherent sources, 
 showing the transition from low noise to high noise regime. 

The perturbative treatment presented here could be easily extended to atoms with more than three levels
interacting with two noisy beams. The analytical solution could be applied for instance, to the five level system 
in \cite{Xiao14_2}. In such a kind of system, magnetic fields are applyied to the Zeeman structure, where the frequency 
domain approach adopted here would show a richer correlation spectrum compared to the time domain approach. Under
certain conditions of magnetic field, the three pairs of sideband resonances could be coupled among them. Such kind 
of system would have applications in a light-matter interface for processing information in continuous variables.

This work was supported by grant \#$2004/13587-0$ and  grant \#$2011/22410-0$, S\~ao Paulo Research Foundation (FAPESP),  CNPq and CAPES (Brazilian agencies), and INCT-IQ (Instituto Nacional de Ci\^encia e Tecnologia de Informa{\c c}\~ao Qu\^antica).
The authors thank Prof. Paulo Nussenzveig and Dr. Ashok Kumar for fruitful discussions
                                                    
\appendix
\section{Bloch equations for a thee level system}\label{apendBloch}
From the hamiltonian (\ref{hamiltonian}) we obtain the following Bloch equations for the rapidly varying variables described by the vector $\mathbf{y}=(\tilde{\rho}_{11},\tilde{\rho}_{22},\tilde{\rho}_{13},\tilde{\rho}_{31},\tilde{\rho}_{23},\tilde{\rho}_{32},\tilde{\rho}_{12},\tilde{\rho}_{21})$, such that:
\begin{align}
\dot{\tilde{\rho}}_{11} = &-i\Omega_1^* \tilde{\rho}_{13}e^{-i(\omega_1 t + \phi_1(t))} + i\Omega_1 \tilde{\rho}_{31}e^{i(\omega_1 t + \phi_1(t))}\nonumber\\&+ \Gamma(1-\tilde{\rho}_{11}-\tilde{\rho}_{22})/2, \label{Rho11}\\
\dot{\tilde{\rho}}_{22} = &-i\Omega_2^* \tilde{\rho}_{23}e^{-i(\omega_2 t + \phi_2(t))} + i\Omega_2 \tilde{\rho}_{32}e^{i(\omega_2 t + \phi_2(t))}\nonumber\\& + \Gamma(1-\tilde{\rho}_{11}-\tilde{\rho}_{22})/2, \label{Rho22}\\
\dot{\tilde{\rho}}_{13} = &  \left(\frac{\Gamma}{2}+i\omega_{31}\right)\tilde{\rho}_{13}+ i\Omega_1 (1-2\tilde{\rho}_{11} -\tilde{\rho}_{22})e^{i(\omega_1 t + \phi_1(t))}\nonumber\\& - i\Omega_2 \tilde{\rho}_{12}e^{i(\omega_2 t + \phi_2)}\label{Rho13}\\  
\dot{\tilde{\rho}}_{23} = &\left(\frac{\Gamma}{2}+i\omega_{32} \right)\tilde{\rho}_{23}+ i\Omega_2 (1-2\tilde{\rho}_{22}-\tilde{\rho}_{11})e^{i(\omega_2 t + \phi_2(t))}\nonumber\\& - i\Omega_1 \tilde{\rho}_{21}e^{i(\omega_1 t + \phi_1(t))} \label{Rho23}\\
\dot{\tilde{\rho}}_{12} = & -(\gamma_d-\omega_{31}+\omega_{32})\tilde{\rho}_{12}+ i\Omega_1 \tilde{\rho}_{32} e^{i(\omega_1 t + \phi_1(t))}\nonumber\\& -i\Omega_2^* \tilde{\rho}_{13}e^{-i(\omega_2 t + \phi_2(t))} \label{Rho12}
\end{align}
where $\omega_{3i}$ correspond to the energy differences between the atomic states, $\Omega_i$ represent the Rabi frequencies, $\Gamma$ is the natural linewidth and $\gamma_d$ is the decoherence rate between the ground states. We have also considered a closed system, which satisfies the normalization $\tilde{\rho}_{11}+\tilde{\rho}_{22}+\tilde{\rho}_{33}=1$.

In order to obtain a linear equation system without explicit time dependence, we define the slowly varying variables
\begin{align}
\rho_{ii}&=\tilde{\rho}_{ii}\\
\rho_{i3}&=\tilde{\rho}_{i3}e^{-i(\omega_i t +\phi_i(t))}\\
\rho_{12}&=\tilde{\rho}_{12}e^{-i((\omega_1-\omega_2) t +\phi_1(t)-\phi_2(t))}
\end{align}
The matrix representation of such transformation is given by
\begin{align}
 \mathbf{x}(t)=e^{-i\textbf{N}_1(\omega_1 t+\phi_1(t))}e^{-i\textbf{N}_2(\omega_2 t+\phi_2(t))}  \mathbf{y}(t) , \label{yxtransformAppend}
\end{align}
and the matrices $\mathbf{N}_1$ and $\mathbf{N}_2$ are defined as 
\begin{align}
\mathbf{N}_1=\begin{bmatrix}
0&0&0&0&0&0&0&0 \\
0&0&0&0&0&0&0&0 \\
0&0&1&0&0&0&0&0 \\
0&0&0&-1&0&0&0&0 \\ 
0&0&0&0&0&0&0&0 \\
0&0&0&0&0&0&0&0 \\
0&0&0&0&0&0&1&0 \\
0&0&0&0&0&0&0&-1\\
\end{bmatrix},\\
\mathbf{N}_2=\begin{bmatrix}
0&0&0&0&0&0&0&0 \\
0&0&0&0&0&0&0&0 \\
0&0&0&0&0&0&0&0 \\
0&0&0&0&0&0&0&0\\ 
0&0&0&0&1&0&0&0\\
0&0&0&0&0&-1&0&0\\
0&0&0&0&0&0&-1&0\\
0&0&0&0&0&0&0&1\\
\end{bmatrix},\label{MatrizN1N2}
\end{align}
where $[\mathbf{N}_1,\mathbf{N}_2]=0$. Therefore, with the vectors defined in eq.~(\ref{yxtransformAppend}) and the matrices in eq.(\ref{MatrizN1N2}), the Bloch equations (\ref{Rho11}-\ref{Rho23}) with their conjugates, can be written as 
\begin{align}
\frac{d\mathbf{y}(t)}{dt}=-e^{i\mathbf{N}_1(\omega_1 t+\phi_1(t))}e^{i\mathbf{N}_2(\omega_2 t+\phi_2(t))} \mathcal{A}_{3N} \mathbf{x}(t) + \mathbf{y}_0, \label{dydtAppend}
\end{align}
where the interaction matrix $\mathcal{A}_{3N}$ is defined as 
\begin{widetext}
\begin{align}
\mathcal{A}_{3N}=\begin{pmatrix}
 \Gamma/2 & \Gamma/2 & i\Omega_1^* & -i\Omega_1 & 0 & 0  & 0 & 0 \\
 \Gamma/2 & \Gamma/2 &0  & 01 & i\Omega_2^* & -i\Omega_2 & 0 & 0 \\
2i\Omega_1 & i\Omega_1 & (\Gamma/2-i\omega_{(31)}) & 0 & 0 & 0 & i\Omega_2 & 0 \\
-2i\Omega_1^* & -i\Omega_1^* & 0 & (\Gamma/2+i\omega_{(31)}) & 0 & 0 & 0 & -i\Omega_2^* \\
i\Omega_2 & 2i\Omega_2& 0 & 0 & (\Gamma/2-i\omega_{(32)}) & 0 & 0 & i\Omega_1^* \\
-i\Omega_2^* & -2i\Omega_2^* & 0 & 0 & 0 & ( \Gamma/2+i\omega_{(32)}) & -i\Omega_1 & 0 \\
0 & 0 & i\Omega_2 & 0 & 0& -i\Omega_1^* & (\gamma_d-i\omega_{(21)} ) & 0 \\
0 & 0 & 0 & -i\Omega_2^* & i\Omega_1 & 0 & 0 & (\gamma_d+i\omega_{(21)}) \\
\end{pmatrix}\label{MatrixA3N}.
\end{align}
\end{widetext}
with $\omega_{(21)}=\omega_{(31)}-\omega_{(32)}$ and 
\begin{align}
\mathbf{y}_0=(\Gamma/2 ,\Gamma/2,-i\Omega_1e^{i(\omega_1t+\phi_1)}, 
i\Omega_1^*e^{-i(\omega_1t+\phi_1)},\nonumber\\
-i\Omega_2 e^{i(\omega_2t+\phi_2)},
i\Omega_2^*e^{-i(\omega_2t+\phi_2)},
0,
0)
\end{align}

\section{Mapping the spectral density with atomic variables}\label{apendA}

With the help of the transformed atomic correlations $\tilde\Pi$ given by eq.~(\ref{Pi_transfm1}), the spectral density matrix can be written as 
\begin{align}
\mathbf{S}(\omega) =&\frac{1}{2\pi}\left[\Lambda_L(\omega) \  \tilde{\Pi}\ \Lambda_R (-\omega)\right]
\end{align}
where we defined
\begin{align}
\Lambda_L(\omega)&=(\mathbf{M}+i\omega)^{-1}\ \mathbf{U}^{-1}\\
\Lambda_R(-\omega)&=\mathbf{U}\ (\mathbf{M}^\dagger - i\omega)^{-1} .
\end{align}
Therefore the matrix elements of the spectral density are
\begin{align}
[\mathbf{S}(\omega)]_{kl} =\sum_{i,j=3}^6 V_{kijl} \  \tilde{\Pi}_{ij} \;, \label{Smatrix_U}
\end{align}
where $V_{kijl}=[\Lambda_L(\omega)]_{ki} [\Lambda_R(-\omega)]_{jl}$.
The sum excludes explicitly the elements with index $i=1,2,7$ and 8 since
\begin{align}
\tilde{\Pi}_{1i}= \tilde{\Pi}_{i1}=0\hspace{1cm}\tilde{\Pi}_{2i}= \tilde{\Pi}_{i2}=0\\
\tilde{\Pi}_{7i}= \tilde{\Pi}_{i7}=0\hspace{1cm}\tilde{\Pi}_{8i}= \tilde{\Pi}_{i8}=0
\end{align}
as a consequence of the matrix elements in eq.~(\ref{sigma_nPert}).

Evaluation of $\tilde\Pi$ stems from the recursive relation of the covariances given by eq.~(\ref{sigma_n}). 
The matrix elements of $\sigma^{(0)}$ represent the the first order atomic response induced by the light-atom  interaction and, 
according to eq.(\ref{sigma_0}) and the transformation (\ref{Pi_transfm1}), it is written as
\begin{widetext}
\begin{align}
\tilde{\sigma}^{(0)}=2
\begin{bmatrix}
0&0&0&0&0&0&0&0\\
0&0&0&0&0&0&0&0\\
0&0&\text{Im}^2 p_1& -\text{Im}\ p_1 \text{Re}\ p_1& \text{Im}\ p_1 \text{Im}\ p_2& -\text{Im}\ p_1 \text{Re}\ p_2&0&0\\
0&0& -\text{Im}\ p_1 \text{Re}\ p_1& \text{Re}^2 p_1&-\text{Im}\ p_2 \text{Re}\ p_1 & \text{Re}\ p_1 \text{Re}\ p_2&0&0\\
0&0&\text{Im}\ p_1 \text{Im}\ p_2& -\text{Im}\ p_2 \text{Re}\ p_1 &\text{Im}^2 p_2&-\text{Im}\ p_2 \text{Re}\ p_2&0&0\\
0&0&-\text{Im}\ p_1 \text{Re}\ p_2& \text{Re}\ p_1 \text{Re}\ p_2&-\text{Im}\ p_2 \text{Re}\ p_2 &\text{Re}^2 p_2&0&0\\
0&0&0&0&0&0&0&0&\\
0&0&0&0&0&0&0&0&\\
\end{bmatrix}.\label{CovHeuristica_trans}
\end{align}
\end{widetext}

We finally have a set of terms with leading contributions from the coherence between levels 1 and 3
  \begin{align}
  \tilde{\Pi}_{33}&=\bar{\gamma} \left( 2\text{Im}^2p_1 + \sum_{n=1}^\infty \bar{\gamma}^n\tilde{\sigma}_{33}^{(2n)}\right),\nonumber\\
    \tilde{\Pi}_{44}&=\bar{\gamma} \left( 2 \text{Re}^2p_1 + \sum_{n=1}^\infty \bar{\gamma}^n\tilde{\sigma}_{44}^{(2n)}\right),\nonumber\\
      \tilde{\Pi}_{34}&=\bar{\gamma} \left( -2\text{Im}p_1\text{Re}p_1 +\sum_{n=1}^\infty \bar{\gamma}^n\tilde{\sigma}_{34}^{(2n)}\right),
\end{align}  
 another set with leading terms related to levels 2 and 3
\begin{align}
 \tilde{\Pi}_{55} &= \bar{\gamma}  \left( 2  \text{Im}^2p_2 + \sum_{n=1}^\infty \bar{\gamma}^n\tilde{\sigma}_{55}^{(2n)}\right),\nonumber\\
\tilde{\Pi}_{66}&=\bar{\gamma}  \left( 2\text{Re}^2p_2 + \sum_{n=1}^\infty \bar{\gamma}^n\tilde{\sigma}_{66}^{(2n)}\right),\nonumber \\
\tilde{\Pi}_{56}&=\bar{\gamma}  \left( -2\text{Im}p_2\text{Re}p_2 +\sum_{n=1}^\infty \bar{\gamma}^n\tilde{\sigma}_{56}^{(2n)}\right),
\end{align}
 and a final set involving crossed terms of both coherences
 \begin{align}
  \tilde{\Pi}_{35}&=\bar{\gamma} \left( 2  \text{Im} p_1\text{Im}p_2 + \sum_{n=1}^\infty \bar{\gamma}^n\tilde{\sigma}_{35}^{(2n)}\right),\nonumber
 \end{align}
 \begin{align}
 \tilde{\Pi}_{46}&=\bar{\gamma} \left( 2 \text{Re}p_1\text{Re}p_2 + \sum_{n=1}^\infty \bar{\gamma}^n\tilde{\sigma}_{46}^{(2n)}\right),\nonumber\\
   \tilde{\Pi}_{36}&=\bar{\gamma} \left( -2\text{Im}p_1\text{Re}p_2 +\sum_{n=1}^\infty \bar{\gamma}^n\tilde{\sigma}_{36}^{(2n)} \right),\nonumber\\
  \tilde{\Pi}_{45}&=\bar{\gamma} \left( -2\text{Im}p_2\text{Re}p_1 +\sum_{n=1}^\infty \bar{\gamma}^n\tilde{\sigma}_{45}^{(2n)} \right).
\end{align}
 
Matrix~(\ref{CovHeuristica_trans}) contains all the necessary elements for calculating the $g^{(2)}(0)$ function of the heuristic model in eq.(\ref{g20}). 
In the frequency domain, the noise spectra in eqs.(\ref{S11})-(\ref{S12}) are given by products of the expansion of the covariances $\tilde\Pi_{ij}$ by weighting factors that depend on the analysis frequency~$\omega$:
\vspace{-.3cm}\begin{align}
S^I(\omega)_{11}& = \sum_{i,j=3}^6 \alpha_{ij}(\omega)\ \tilde{\Pi}_{ij}\\
S^I(\omega)_{22}& = \sum_{i,j=3}^6 \beta_{ij}(\omega)\ \tilde{\Pi}_{ij}\;,\\
S^I(\omega)_{12}& = \sum_{i,j=3}^6 \nu_{ij}(\omega) \tilde{\Pi}_{ij}
\end{align}
where
\begin{align}
\alpha_{ij}(\omega)=&[V_{3ij3}(\omega)+ V_{4ij4}(\omega) -V_{3ij4}(\omega)-V_{4ij3}(\omega)]\label{alpha_ij}
\end{align}
\vspace{-.2cm}\begin{align}
\beta_{ij}(\omega)=&[V_{5ij5}(\omega)+ V_{6ij6}(\omega) -V_{5ij6}(\omega)-V_{6ij5}(\omega)]\label{beta_ij}\;.
\end{align}
\vspace{-.2cm}\begin{align}
\nu_{ij}(\omega)=&1/2[V_{3ij6}(\omega)+ V_{4ij5}(\omega) -V_{3ij5}(\omega)-V_{4ij6}(\omega)\label{nu_ij}\nonumber\\
 &+V_{6ij3}(\omega)- V_{6ij4}(\omega) +V_{5ij4}(\omega)-V_{5ij3}(\omega)]
\end{align}

Noise power spectra are now evaluated for these expanded terms as
 \begin{align}
S^I_{11}(\omega) &=
\alpha_{33}(\omega)  \tilde{\Pi}_{33}
+\alpha_{44}(\omega)  \tilde{\Pi}_{44+}+C_{11}(\omega), \label{s11_pert_final}
\end{align}
where the complementary coefficient $C_{11}$ is defined 
\begin{align}
C_{11}(\omega)&= \tilde{\alpha}_{34}  \tilde{\Pi}_{34}+\alpha_{55}(\omega)\tilde{\Pi}_{55} +\alpha_{66}(\omega)\tilde{\Pi}_{66}\nonumber \\
&+\tilde{\alpha}_{56}(\omega)\tilde{\Pi}_{56} + \tilde{\alpha}_{35}(\omega)\tilde{\Pi}_{35} +\tilde{\alpha}_{46}(\omega)\tilde{\Pi}_{46}\nonumber\\
&+ \tilde{\alpha}_{36}(\omega)\tilde{\Pi}_{36}+\tilde{\alpha}_{45}(\omega)\tilde{\Pi}_{45}, \label{C_11}
\end{align}
with $\tilde{\alpha}_{ij}(\omega)=\alpha_{ij}(\omega) + \alpha_{ji}(\omega)$. These terms
are related to contributions from the polarization of the other transition, and cross terms. 
Their role is discussed in the main text.

Similarly, we have for the noise spectra for the second field
\begin{align}
S^I_{22}(\omega) &=
  \beta_{55}(\omega)  \tilde{\Pi}_{55}+\beta_{66}(\omega)  \tilde{\Pi}_{66}+C_{22}(\omega), \label{s22_pert_final}
\end{align}
and complementary coefficients 
\begin{align}
C_{22}(\omega)&= \tilde{\beta}_{56}  \tilde{\Pi}_{56} +\beta_{33}(\omega)\tilde{\Pi}_{33} +\beta_{44}(\omega)\tilde{\Pi}_{44}\nonumber\\
&+\tilde{\beta}_{34}(\omega)\tilde{\Pi}_{34}+ \tilde{\beta}_{35}(\omega)\tilde{\Pi}_{35} +\tilde{\beta}_{46}(\omega)\tilde{\Pi}_{46}\nonumber\\
&+ \tilde{\beta}_{36}(\omega)\tilde{\Pi}_{36}+\tilde{\beta}_{45}(\omega)\tilde{\Pi}_{45},\label{C_22}
\end{align}
with $\tilde{\beta}_{ij}(\omega)=\beta_{ij}(\omega) + \beta_{ji}(\omega)$.\\

And, finally, for the correlation
\begin{align}
S^I_{12}(\omega) &=
\tilde{\nu}_{35}(\omega)\tilde{\Pi}_{35}+\tilde{\nu}_{46}(\omega)\tilde{\Pi}_{46}\nonumber\\
&+\tilde{\nu}_{36}(\omega)\tilde{\Pi}_{36}+\tilde{\nu}_{45}(\omega)\tilde{\Pi}_{45}+C_{12}(\omega),\label{s12_pert_final}
\end{align}
with $\tilde{\nu}_{ij}(\omega)=\nu_{ij}(\omega) + \nu_{ji}(\omega) $ and complementary coefficients
\begin{align}
C_{12}(\omega) &=  \nu_{33}(\omega)\tilde{\Pi}_{33} +\nu_{44}(\omega)\tilde{\Pi}_{44} +\nu{\beta}_{34}(\omega)\tilde{\Pi}_{34}\nonumber\\
 &+\nu_{55}(\omega)\tilde{\Pi}_{55} +\nu_{66}(\omega)\tilde{\Pi}_{66} +\nu{\alpha}_{56}(\omega\tilde{\Pi}_{56}.\label{C_12}
\end{align}

It is worth noting that elements $  \tilde{\Pi}_{33}$,  $  \tilde{\Pi}_{44}$, and  $  \tilde{\Pi}_{34}$ describe the atomic response with respect to the light beam $\mathbf{E}_1$ since they are proportional to the absorption $\text{Im}\ p_1$ and dispersion $\text{Re}\ p_1$. The elements $ \tilde{\Pi}_{55}$,  $  \tilde{\Pi}_{66}$ and  $  \tilde{\Pi}_{56}$ represent the atomic response with respect to field $\mathbf{E}_2$ according to its dependence on  $\text{Im}\ p_2$ and $\text{Re}\ p_2$. Finally, the cross terms $  \tilde{\Pi}_{35}$,  $ \tilde{\Pi}_{46}$ ,  $  \tilde{\Pi}_{36}$ and   $  \tilde{\Pi}_{45}$  describe the main elements that contribute to the correlation $C(\omega)$.

\end{document}